\documentclass{article}
\usepackage[utf8]{inputenc}
\usepackage{graphicx}
\usepackage{siunitx}
\usepackage{amsmath, amssymb}
\usepackage{xcolor}
\usepackage{authblk}
\usepackage{url}
\DeclareGraphicsExtensions{png, svg}
\DeclareSIUnit\dBm{dBm}

\def\modif#1{\textcolor{black}{#1}}

\title{Ground State Cooling of an Ultracoherent Electromechanical System}

\author[1,2]{Yannick Seis}
\author[1,2]{Thibault Capelle}
\author[1,2]{Eric Langman}
\author[1,2]{Sampo Saarinen}
\author[1,2]{Eric Planz}
\author[1,2,*]{Albert Schliesser}

\affil[1]{Niels Bohr Institute, University of Copenhagen, Blegdamsvej 17, 2100 Copenhagen, Denmark}
\affil[2]{Center for Hybrid Quantum Networks (Hy-Q), Niels Bohr Institute, University of Copenhagen, Copenhagen, Denmark}
\affil[*]{email: albert.schliesser@nbi.dk}

\date{}

\begin{document}

\maketitle

\begin{abstract}

Cavity electromechanics relies on parametric coupling between microwave and mechanical modes to manipulate the mechanical quantum state, and provide a coherent interface between different parts of hybrid quantum systems. 
High coherence of the mechanical mode is of key importance in such applications, in order to protect the quantum states it hosts from thermal decoherence. 
Here, we introduce an electromechanical system based around a soft-clamped mechanical resonator with an extremely high Q-factor ($>10^9$) held at very low (30 mK) temperatures. 
This ultracoherent mechanical resonator is capacitively coupled to a microwave mode, strong enough to enable ground-state-cooling of the mechanics ($\bar n_\mathrm{min}=0{.}76\pm 0{.}16$). 
This paves the way towards exploiting the extremely long coherence times ($t_\mathrm{coh}>100\,\mathrm{ms}$) offered by such systems for quantum information processing and state conversion. 

\end{abstract}

\section{Introduction}

The field of cavity electromechanics \cite{Lehnert2014, Clerk2020} investigates mechanical resonators which are parametrically coupled to radio-frequency or microwave circuits. Analogous to cavity optomechanics \cite{Aspelmeyer2014}, this coupling is at the heart of a broad set of phenomena and techniques of interest in quantum science and technology. 
They range from ground-state cooling of the mechanics \cite{Teufel2011, Yuan2015, Noguchi2016}, via entanglement and squeezing \cite{Palomaki2013b, Wollman2015, Ockeloen2018, Delaney2019}, to coherent microwave-optical \cite{Andrews2014, Bagci2014} (see also \cite{Midolo2018} and references therein) and superconducting qubit-mechanical interfaces \cite{Chu2017, Arrangoiz2019, Bienfait2019, Mirhosseini2020}. 

For most of these applications, a long \modif{coherence time}
\begin{equation}
t_\mathrm{coh}=\frac{\hbar Q}{k_\mathrm{B} T_\mathrm{bath}}= \frac{1}{\bar{n}_\mathrm{th}\Gamma_\mathrm{m}},
\label{eq:coh}
\end{equation}
of the mechanical system is favourable. Here, $Q=\Omega_\mathrm{m}/\Gamma_\mathrm{m}$ is the mechanical quality factor defined as the ratio of the mechanical (angular) frequency $\Omega_\mathrm{m}$ and its energy decay rate $\Gamma_\mathrm{m}$; $T_\mathrm{bath}$ the resonator's bath temperature; $\hbar$ and $k_\mathrm{B}$ the reduced Planck and the Boltzmann constant, respectively; and 
$\bar{n}_\mathrm{th}\approx k_\mathrm{B}T_\mathrm{bath}/\hbar \Omega_\mathrm{m}$ is the equivalent occupation of the thermal bath.

For state-of-the-art electromechanical systems operated at milliKelvin temperatures, typical Q-factors are $\lesssim 10^7$ and coherence times are at most 1 millisecond.
This applies to a wide variety of systems, including aluminum vacuum gap capacitors \cite{Palomaki2013,Palomaki2013b,Weinstein2014,Wollman2015,Ockeloen2018,Delaney2019}, metallized silicon nitride membranes \cite{Noguchi2016,Andrews2014, Higginbotham2018,Delaney2021} and strings \cite{Zhou2013}, quantum acoustic devices \cite{Chu2017, Bienfait2019}, as well as piezoelectrically coupled nanophononic crystals \cite{Arrangoiz2019,Mirhosseini2020}.
As a notable exception, $Q\approx 10^8$ has been reported for a metallized silicon nitride membrane in 2015 \cite{Yuan2015APL}. 
Its enhanced performance over similar devices \cite{Noguchi2016,Andrews2014, Higginbotham2018,Delaney2021} might be linked to its particularly low operation temperature ($\sim 10\,\mathrm{mK}$) and frequency ($\sim 100\,\mathrm{kHz}$)--which, among other things, can make operation in the simultaneously overcoupled and resolved-sideband regime challenging.

On the other hand, recent progress in the design of mechanical systems has allowed reaching quality factors in excess of $10^9$ at Mega- to Gigahertz frequencies \cite{Tobar2014, Tsaturyan2017, Rossi2018, Painter2020, Beccari2021a, Beccari2021b}.
 At milliKelvin temperatures, such ultracoherent mechanical devices can reach $t_\mathrm{coh}>100\ \mathrm{ms}$, some two orders of magnitude beyond the typical performance of state-of-the-art devices (provided excess dephasing \cite{Painter2020} is not an issue). 
However, so far, the mechanics' coupling to microwave modes has either been extremely weak \cite{Tobar2014}, or absent because of lacking functionalization through e.g. metallization \cite{Tsaturyan2017, Rossi2018,Painter2020, Beccari2021a, Beccari2021b}.
For this reason, these mechanical systems could not yet be harnessed in electromechanics.

Here, we realize an ultracoherent electromechanical system based on a soft-clamped silicon nitride membrane \cite{Tsaturyan2017}.
Following earlier work \cite{Andrews2014,Yuan2015,Yuan2015APL, Noguchi2016}, we functionalize it with a superconducting metal pad. 
This allows coupling it to a microwave resonator to implement the standard optomechanical Hamiltonian
\begin{equation}
    \hat{H}_\mathrm{int}=\hbar g_0\hat{a}^\dagger\hat{a}\left(\hat{b}+\hat{b}^\dagger\right),
\end{equation}
as shown in previous works \cite{Lehnert2014}.
Here, $g_0/2\pi$ is the microwave frequency shift due to the zero point fluctuation of the mechanical resonator, $\hat{a}$($\hat{b}$) is the photon (phonon) annihilation operator. Under a strong pump, the system is populated by a mean coherent field around which the Hamiltonian can be linearized to
\begin{equation}
    \hat{H}_\mathrm{int}\approx\hbar g_0\sqrt{n}\left(\delta\hat{a}^\dagger+\delta\hat{a}\right)\left(\delta\hat{b}+\delta\hat{b}^\dagger\right),
\end{equation}
where $n$ is the mean photon number in the cavity, and the annihilation operators are here small displacements around a mean coherent field. In this case, well-established concepts and methods of optomechanics as described e.g.\ in \cite{Aspelmeyer2014} apply.
In our work, we realize sufficient coupling strength to cool the mechanical mode to its quantum mechanical ground state.
This implies that we have achieved a quantum cooperativity $C_\mathrm{q}>1$ \cite{Aspelmeyer2014} and heralds the possibility to deploy soft-clamped mechanical resonators for applications in quantum electromechanics.

\section{Results}
\subsection{Electromechanical system}

\begin{figure}[htbp]
    \centering
    \includegraphics[width=.9\linewidth]{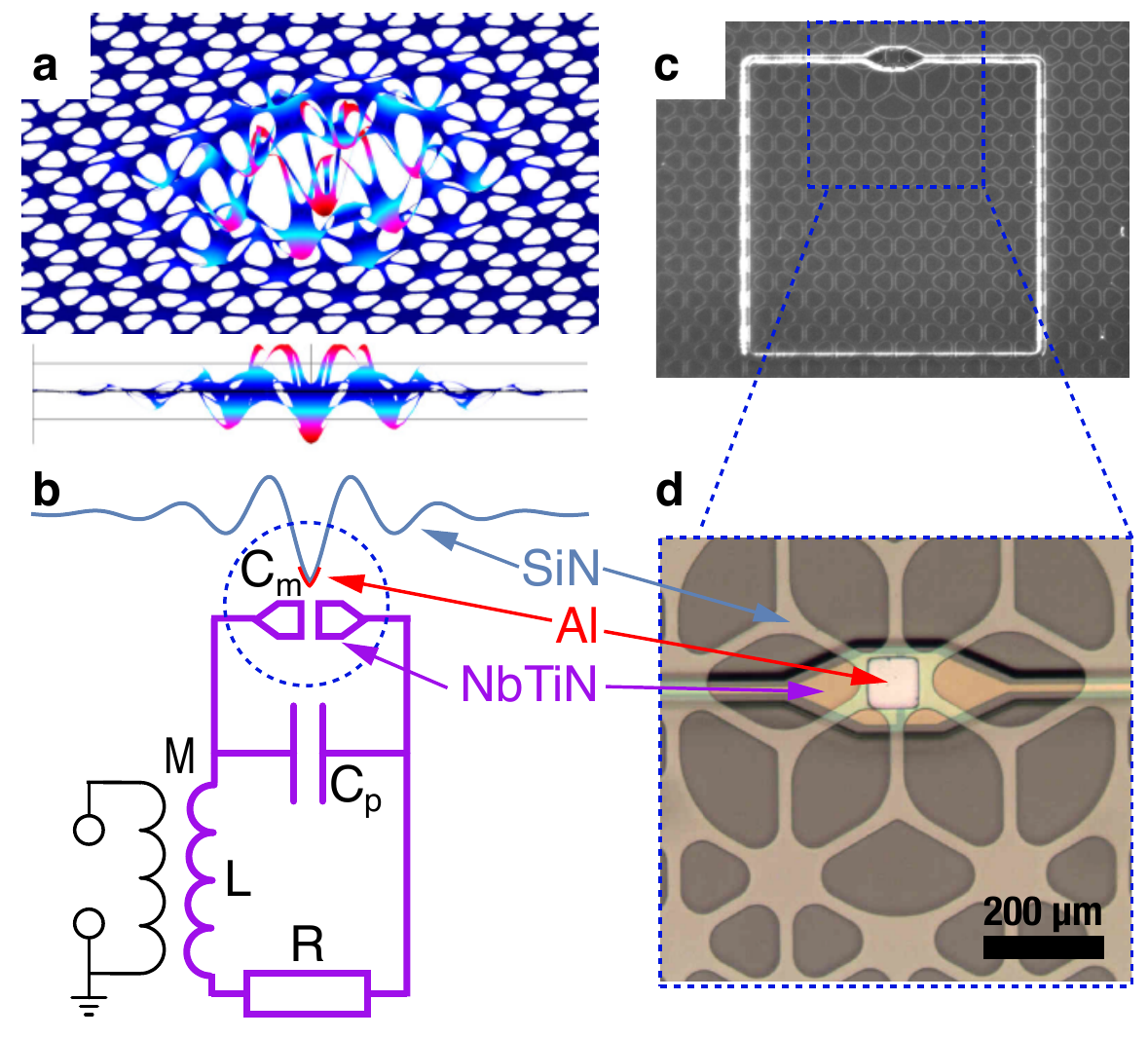}
    \caption{\label{fig:sample}
    Electromechanical system. 
    (a) Bird's (top) and side (bottom) view of the simulated displacement of the mechanical mode localized at the defect in the phononic crystal patterned into a  silicon nitride (SiN) membrane. False-color indicates displacement amplitude from small (blue) to large (red).
    (b) The membrane defect is metallized with a pad of aluminum (Al) and brought into proximity of two electrode pads on a different chip, thereby forming a mechanically compliant capacitance $C_\mathrm{m}$.
    This capacitor is part of a microwave `loop-gap' resonator made from the superconductor NbTiN, together with a parallel parasitic capacitance $C_\mathrm{p}$, inductivity $L$ and resistance $R$. 
    Microwave power is coupled into this circuit through the mutual inductance $M$. 
    (c) Gray-scale optical micrograph (top view) of the sandwich, in which the microwave loop-gap resonator (bright square) shines through the largely transparent patterned membrane.
    (d)~Color zoom onto the mechanically compliant capacitor, showing the square Al metallization on the patterned membrane above the NbTiN capacitor pads.
    }
\end{figure}

The system studied here in shown in Fig.~\ref{fig:sample}.
It consists of a 63-nm thick soft-clamped membrane made of silicon nitride \cite{Tsaturyan2017}.
A square portion of its central defect (an area of approximately $60 \times 60~\mathrm{\mu m}^2$) is covered with a 50-nm thick layer of aluminum. 
This superconducting pad is placed, using a flip-chip assembly, closely above the capacitive electrodes of a planar loop gap resonator fabricated from a 100-nm thick layer of NbTiN, forming a resonant LC circuit.
The motion of the metalized membrane modulates the capacitance and in turn the resonance frequency of the microwave circuit, thereby forming a canonical electromechanical system \cite{Yuan2015,Andrews2014, Noguchi2016}. 

The device is read out by inductive coupling to a coaxial transmission line and is placed on a mechanical damper, for vibration isolation \cite{Wit2019}, mounted on the mixing chamber plate of a dilution refrigerator (see Methods for details).

From microwave reflection measurements performed by the vector network analyser, we extract a cavity resonance frequency $\omega_\mathrm{c}/2\pi = \SI{8.349}{\GHz}$, a total linewidth $\kappa /2\pi = \SI{240}{\kHz}$ and an outcoupling efficency $\eta=\kappa_\mathrm{ex}/\kappa \sim 0.8$. With a mechanical mode at $\Omega_\mathrm{m}/2\pi = \SI{1.486}{\MHz}$, the system is well sideband-resolved ($\kappa \ll \Omega_\mathrm{m}$).

\begin{figure}[htbp]
    \centering
    \includegraphics[width=\linewidth]{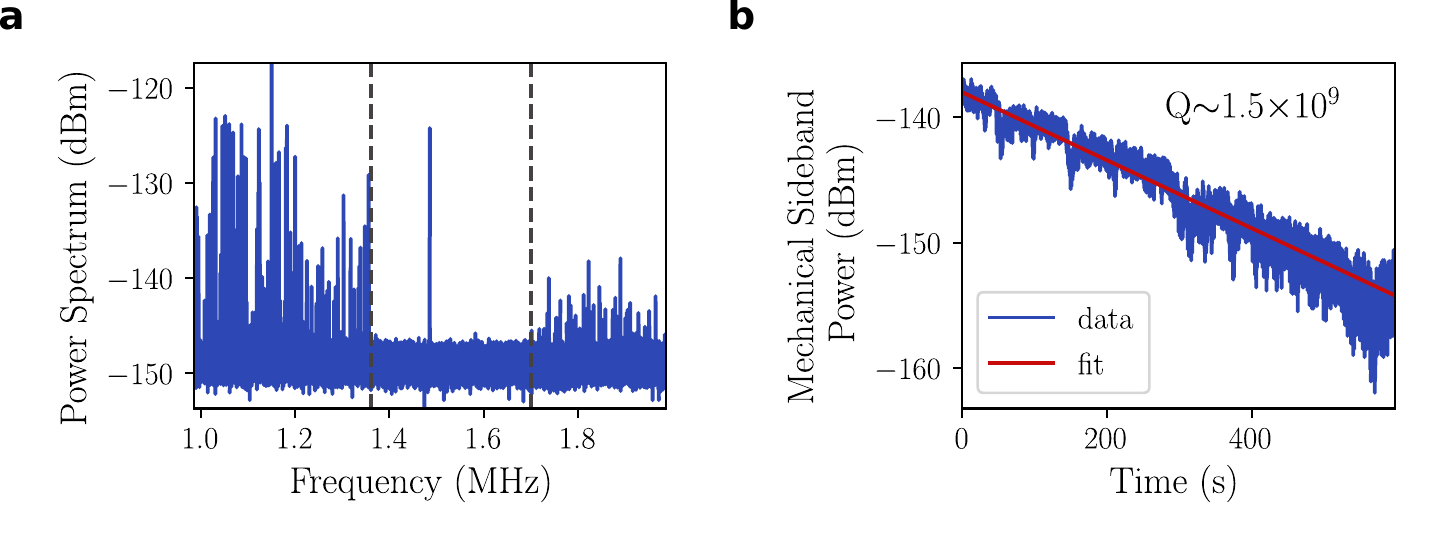}
    \caption{\label{fig:mechanics}
   Mechanical properties. (a) Thermal noise spectrum showing a large bandgap (whose limits are indicated with dashed lines) around the mode of interest at $\sim 1.5\,\mathrm{MHz}$. (b) Mechanical ringdown of the defect mode of a metallized membrane at cryogenic temperature, yielding an ultra-high quality factor.
    }
\end{figure}

The soft-clamped membranes utilized in this work represent a new design of phononic membrane resonators, one which we find to have superior characteristics for electromechanical functionalization. Each membrane of this new design is referred to as a `Lotus,' inspired by the resemblance of the defect-defining perforations to the large petals of various species of lotus flowers. Not only do we observe that Lotus-class designs possess  larger bandgaps, they are capable of localizing a single out-of-plane mechanical mode centered in that enlarged bandgap, with maximum amplitude at the center of the defect  Importantly, this single mode remains well-isolated from the bandgap edges after aluminum metallization, as shown in Fig.~\ref{fig:mechanics}.  Finally, we find such metallized lotuses are able to yield ultrahigh mechanical quality factors in excess of $10^9$ at cryogenic temperature, as measured by energy ringdown (see Fig.~\ref{fig:mechanics}).

\subsection{Calibrations}
\label{sec:calibrations}

\begin{figure}
    \centering
    \includegraphics[width=1.\linewidth]{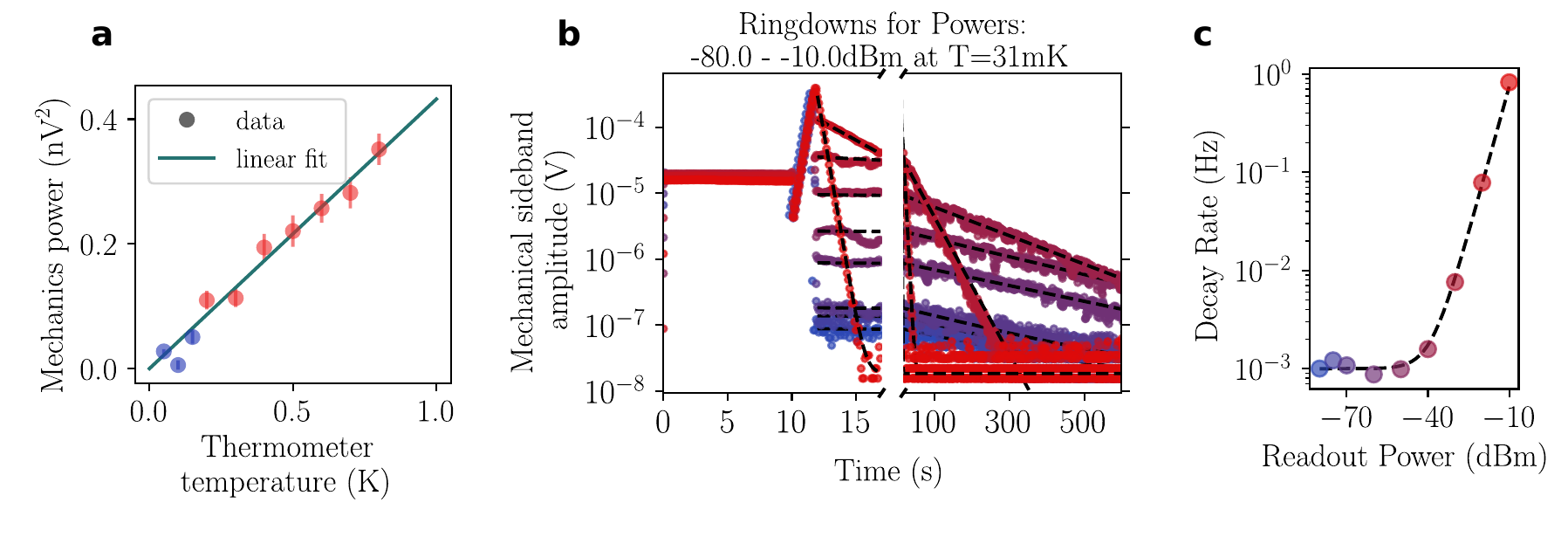}
    \caption{\label{fig:calibration}Electro-mechanical calibration: 
    (a) The mechanical occupation is calibrated by thermal anchoring at temperatures above \SI{200}{\milli \K}. The linear relationship between the area of the mechanical peak in the spectrum and the sample holder temperature confirms the mechanics is thermalized. Only the red data are used for the fit (See main text). \modif{Error bars are std. dev. of the mechanical sideband area fits}. 
    (b) A mechanical energy ringdown series measured as function of the applied cooling power, measured at \SI{30}{\milli \K}. Overlayed temporal series show repeatable initialization of the mechanical energy (up to \SI{\sim 12}{\s}) and increasing decay rates as the cooling power is turned up. 
    (c) The fit of mechanical decay rates gives the intrinsic decay rate $\Gamma_\mathrm{m}$, without dynamical backaction, and the corner power $P_0$, where the cooling rate $\Gamma_\mathrm{e}(P_0)$ is equal to $\Gamma_\mathrm{m}$. Points' color code is the same as in panel B.}
    \label{fig:calib}
\end{figure}

We establish the phonon occupation of the mechanical resonator via its equilibration to the controlled thermal bath provided in the cryostat.
We drive the electromechanical system with a tone at frequency $ \omega_\mathrm{p}$, red-detuned from the cavity resonance $\Delta\equiv \omega_\mathrm{p}-\omega_\mathrm{c}\approx -\Omega_\mathrm{m}<0 $ at a fixed, low  power (\SI{-45}{\dBm} at the source) such that dynamical backaction \cite{Aspelmeyer2014} is negligible. 
Then, after further amplification and carrier cancellation (see Methods), we measure the spectral area occupied by the mechanical sideband (i.e.\ the total microwave power) around the frequency $\omega_\mathrm{p}+\Omega_\mathrm{m}$ for a range of sample temperatures as measured by the cryostat thermometer. 
At temperatures above \SI{\sim 200}{\milli \K} (see Fig.~\ref{fig:calib}A), we observe a linear relationship between temperature and mechanical sideband area. 
This proportionality is interpreted as the sample being in a thermal equilibrium with the mixing chamber plate. 
Using Bose-Einstein statistics for thermal states $\bar{n}_\mathrm{th}=(e^{\hbar\Omega_\mathrm{m}/k_\mathrm{B}T_\mathrm{bath}}-1)^{-1}$, we extract a calibration constant between mechanical sideband area and mechanical occupation in quanta. At temperatures below \SI{\sim 200}{\milli \K}, dynamical backaction is small but nonnegligible ($\lesssim 15\%$). This has been corrected for, together with the temperature-dependent microwave and mechanical damping (see supplementary material). Figure \ref{fig:calib}A  shows the resulting thermalization of the mechanical oscillator to the base plate of the cryostat. 
From this analysis, we infer that at the lowest cryostat temperature of 30~mK, the mechanical mode is coupled to a bath at $T_\mathrm{bath}\approx 80\,\mathrm{mK}.$ 

Next, we calibrate the dynamical backaction by  performing mechanical ringdown measurements under red-detuned microwave drives with varying powers.
Figure \ref{fig:calib}B shows superimposed ringdown sequences under increasing microwave power.
For these measurements, we initialize the mechanics into a large coherent state by phase modulation of the red-detuned pump (duration \SI{10}{\s}), then amplify this coherent state by placing the pump on the blue side of the cavity (duration \SI{1.75}{\s}).
Finally, we let the mechanics ring down for \SI{600}{\s} with the red-detuned microwave pump at power $P$ varying from \SIrange{-80}{-10}{\dBm} (at the output of the signal generator).
We fit the ringdowns to exponential decays where the time constants are the inverse angular decay rates $\Gamma_\mathrm{eff}^{-1}$. 

The resulting decay rates as function of pump power are shown in Fig.~\ref{fig:calib}C, together with a fit using the model
\begin{align}
    \Gamma_\mathrm{eff}(P) = \Gamma_\mathrm{m} + \Gamma_\mathrm{e}(P) = \Gamma_\mathrm{m} \left( 1 + \frac{P}{P_\mathrm{0}}\right).
\end{align}

Here, $\Gamma_\mathrm{m}$ is the intrinsic loss rate of the mechanical resonator, while $\Gamma_\mathrm{e}(P)$ is the damping imparted by the dynamical backaction of the microwave mode \cite{Aspelmeyer2014}.
We introduce $P_\mathrm{0}$ as the  power at which the pump-induced decay is equal to the intrinsic decay rate $\Gamma_\mathrm{m}$. Note that $P_\mathrm{0}$ depends on the cavity lineshape and the pump detuning from cavity resonance. 
At a cryostat temperature of \SI{30}{\milli \K},  we extract $\Gamma_\mathrm{m}/2\pi = \SI{1.0}{\milli \Hz}$, a quality factor $Q_\mathrm{m} = \Omega_\mathrm{m}/\Gamma_\mathrm{m} = 1.5 \cdot 10^9$ and $P_\mathrm{0} = \SI{-38.7}{\dBm}$ from the dataset shown in Fig.~\ref{fig:calib}. We note that the quality factor is dependent on the sample temperature (see \modif{Supplementary} for a systematic analysis).
We perform our ground state cooling at the same temperature, allowing us to use the $ \Gamma_\mathrm{m}$ and $\Gamma_\mathrm{e}(P)$ obtained from this fit as fixed parameters in all further analysis.

Finally, from a standard calibration technique detailed in Ref.~\cite{Gorodetksy2010}, we extract a single-photon coupling rate $g_\mathrm{0}/2\pi = \SI[separate-uncertainty=true]{0.89 \pm 0.11}{\Hz}$.
From the same calibration, we obtain an overall electronic gain between the sample and the spectrum analyzer. 
From an complementary measurement of transmission of the entire setup, we can infer an attenuation of $(66.5\pm 1)$ dB between the signal source and the sample.
This means that for the highest source power of \SI{+10}{\dBm}, the power at the device input is \SI{-56.5}{\dBm} and the cavity is populated with $3.3 \cdot 10^7$ microwave photons.

\subsection{Ground state cooling}

To reduce the mechanical occupation of our mechanical resonator, we place a strong coherent pump on the red sideband of the microwave cavity
($\Delta=-\Omega_\mathrm{m}$), in the same experimental conditions with which we calibrated the phonon occupation (see Section \ref{sec:calibrations}).

In the resolved-sideband limit ($\kappa\ll\Omega_\mathrm{m}$), which we reach in this experiment, and close to the cavity frequency ($|\omega-\omega_\mathrm{c}|\ll\kappa$), the microwave power spectral density in units of noise quanta is then:
\begin{equation}
\label{eq:output}
    S[\omega]=n_\mathrm{add}+4\eta\left(\tilde{n}+1/2\right)+\eta \Gamma_\mathrm{m}\Gamma_\mathrm{e}\frac{\bar{n}_\mathrm{th}+\frac{1}{2}-\left(2+\frac{\Gamma_\mathrm{e}}{\Gamma_\mathrm{m}}\right)\left(\tilde{n}+\frac{1}{2}\right)}{\left(\Gamma_\mathrm{m}+\Gamma_\mathrm{e}\right)^2/4+\left(\omega-\omega_\mathrm{p}-\Omega_\mathrm{eff}\right)^2},
\end{equation}
where we have defined $\tilde{n}=\eta n_c+\left(1-\eta\right)n_0$, with $\eta=\kappa_\mathrm{c}/\kappa$, $\kappa_c$ the coupling rate to the microwave cavity, $n_c$ the microwave noise occupation coming from either the pump phase noise or the cavity frequency noise, $n_0$ the noise occupation of the microwave thermal environment (which is negligible in the considered experimental conditions). In the above expression, $\Omega_\mathrm{eff}/2\pi$ is the effective mechanical frequency including the frequency shift induced by the dynamical backaction.
The measured microwave spectrum is composed of three parts: the background noise $n_\mathrm{add}$, which is due to the HEMT amplifier, the  microwave noise coming from the cavity, which is a Lorentzian whose width is the microwave loss rate $\kappa$, and the mechanical noise transduced into microwave noise \emph{via} the electromechanical coupling. This mechanical feature is a Lorentzian of width $\Gamma_\mathrm{eff}=\Gamma_\mathrm{m}+\Gamma_\mathrm{e}$. 

\begin{figure}
    \centering
    \includegraphics[width=1.\linewidth]{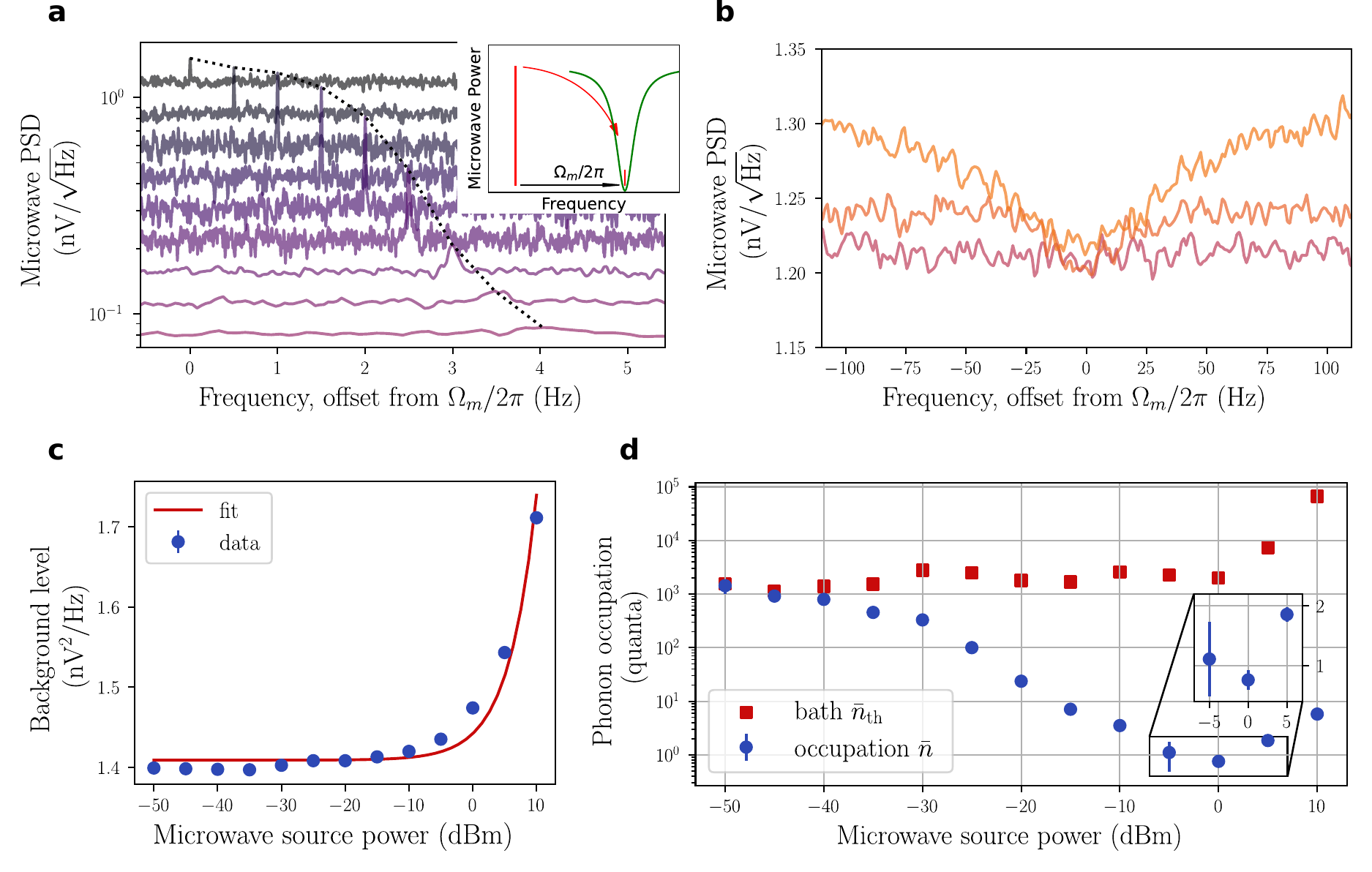}
    \caption{Sideband cooling of the mechanics to its motional ground state. 
    (a) Mechanical power spectral density (PSD) around the defect mode as the cooling power is increased: the peak first increases in height with measurement gain, then visibly broadens as cooling takes place. Spectra are offset downwards and to the right in equal steps respectively as the power is increased. The dotted line is a guide to the eye for the location of the mechanical peaks. \modif{The inset depicts the scattering process due to the mechanics, which creates a sideband (small vertical red line) in the microwave cavity (green dip) at an angular frequency $\Omega_m$ above the pump tone (tall vertical red line).} 
    (b) At high pump powers, the cavity is populated due to the microwave phase noise: the noise occupancy is visible in the `squashing' of the mechanical feature. 
    (c) The increased background level of the mechanical spectrum allows to extract the cavity occupation $\tilde{n}$. 
    (d) The mechanical occupation, calibrated in number of motional quanta, reaches below one phonon by dynamical backaction cooling before being heated up by cavity occupation due to source phase noise. \modif{Error bars are std. dev. of the mechanical sideband area fits.}}
    \label{fig:gsc}
\end{figure}

At low cooperativity $C\approx\Gamma_\mathrm{e}/\Gamma_\mathrm{m}\ll 1$, the signal, divided by the electromechanical gain $\eta\Gamma_\mathrm{e}/\Gamma_\mathrm{m}$ is simply a Lorentzian whose area is proportional to the mechanical bath occupation $\bar{n}_\mathrm{th}\gg \tilde n$. This is the regime where we performed the calibrations presented in section \ref{sec:calibrations}.

At a higher cooperativity $C\approx\Gamma_\mathrm{e}/\Gamma_\mathrm{m}\gg 1$, the rate at which phonons are extracted from the resonators initially exceeds the rate at which new phonons are entering the resonator \emph{via} the mechanical thermal bath. A new equilibrium is established at a reduced temperature of the mechanical resonator,  corresponding to a reduced effective occupation $\bar{n}<\bar{n}_\mathrm{th}$. This appears as a decrease of the area under the mechanical spectrum.

At the highest cooperativities, the microwave noise starts to play a significant role. It originates either from the phase noise of the microwave source or from a cavity frequency noise. We see from equation~(\ref{eq:output}) that the observed signal can then be a \emph{negative} Lorentzian. This does not mean that the temperature of the mechanical mode is negative, but rather that the cross spectrum between the microwave noise in the cavity and the microwave noise transduced to mechanical noise changes the shape of the resulting signal \cite{Weinstein2014}. In this case, inference of the mechanical mode temperature requires the knowledge of the phase noise, which is given by the background of the signal (for $\Gamma_\mathrm{m}+\Gamma_\mathrm{e}\ll|\omega-\omega_\mathrm{c}|\ll\kappa$)\cite{SafaviNaeini2013}:
\begin{equation}
    S[\omega]_{\Gamma_\mathrm{m}+\Gamma_\mathrm{e}\ll|\omega-\omega_\mathrm{c}|\ll\kappa}=\mathcal{A}+\alpha P,
\end{equation}
where $P$ is the pump power, $\alpha P=4\eta^2n_c\approx 4\eta\tilde{n}$, and $\mathcal{A}=n_\mathrm{add}+2\eta+4\eta\left(1-\eta\right)n_0\approx n_\mathrm{add}$.
By fitting the model of Eq.~(\ref{eq:output}) to the experimental spectra comprising both the mechanical feature and the background level, we obtain the parameters $\bar{n}_\mathrm{th}$ and $\tilde{n}$, respectively, at each power level (see Fig. \ref{fig:gsc}). We can then compute the mechanical occupation as \cite{Yuan2015}:
\begin{equation}
    \bar{n}=\frac{\Gamma_\mathrm{m}}{\Gamma_\mathrm{m}+\Gamma_\mathrm{e}}\bar{n}_\mathrm{th}+\frac{\Gamma_\mathrm{e}}{\Gamma_\mathrm{m}+\Gamma_\mathrm{e}}\tilde{n}.
\end{equation}

The minimum inferred occupation is $\bar{n}_\mathrm{min}=0.76\pm 0.16$. This final value is limited by the efficiency of the vibration isolation, which increases the mechanical bath temperature above the thermodynamic temperature (see Methods), and the microwave phase noise at the input of the system. Although an increase of the mechanical bath temperature can be observed for pump powers $\geq 0$ dBm, the contribution of the phase noise is still dominant. Placing a microwave cavity filter at the output of the signal generator, absorbing pump phase noise around the electromechanical cavity resonance, did not improve the result. This suggests that the phase noise is limited by cavity noise rather than the phase noise of the microwave source.

\section{Discussion}

The mechanical occupation calibrated in Fig.~\ref{fig:calib}A along with the measured intrinsic mechanical decay rate furthermore allow us to estimate the mechanics' quantum coherence time.
Following eq.~(\ref{eq:coh}), we extract $t_\mathrm{coh} \approx \SI{140}{\ms}$. 
This is three orders of magnitude larger than for state-of-the-art electromechanical systems \cite{Chu2017,Delaney2019}.
However, further work will be needed to fully confirm the coherence of the mechanical system, ruling out e.~g.~excess decoherence by dephasing \cite{Painter2020}.

At the highest input powers ($P=10\,\mathrm{dBm}$), we achieve a cooperativity $C=\mathcal{O}(10^5)$ and an electromechanical damping on the order of $\Gamma_\mathrm{e}/2\pi\sim 80\, \mathrm{Hz}$.

However, we estimate that the single-photon coupling rate $g_0$ might be increased by an order of magnitude by adjustment of the geometry, in particular the gap between the membrane electrode and its counterelectrodes.
This would immediately boost the coupling (with $\Gamma_\mathrm{e}\propto g_0^2$) and simultaneously alleviate the issues with  microwave phase noise.
Indeed, $g_0/2\pi=7\,\mathrm{Hz}$ and coupling rates well above \SI{100}{\kHz} have been demonstrated in a similar system \cite{Noguchi2016}.
The challenge in transferring this result to our system lies in realizing similarly small capacitive gaps in spite of a significantly larger membrane size, posing stringent requirements on wafer flat- and cleanliness.

Potential applications of the platform introduced here include quantum memories for microwave quantum states \cite{Palomaki2013}, where they could replace or supplement less coherent ($t_\mathrm{coh}^\mathrm{MW}\sim 10\,\mathrm{ms}$), much more bulky microwave resonators \cite{Reagor2013}.
By combining this with an \emph{opto}-mechanical interface \cite{Rossi2018}, e.~g.~by introducing a second defect in the phononic crystal \cite{Catalini2020}, such systems could form part of an electro-opto-mechanical transducer \cite{Andrews2014, Bagci2014}.
One of its key figures of merit, namely the number of added noise quanta, falls proportionally with the coherence time of the mechanics \cite{Zeuthen2020}.
Furthermore, the high mechanical coherence immediately translates to an outstanding force sensitivity. 
This allows for the microwave mode to be used as a sensitive transducer for the motion induced by the physical system of interest, which could be anything from spins \cite{Fischer2019,Kosata2020,Halg2021} to dark matter \cite{Carney2021}.
Nominally, the resonant force noise spectral density of the presented device is $S_{FF}^{1/2}=(2 m \Gamma_\mathrm{m} k_\mathrm{B} T)^{1/2} \approx 650\,\mathrm{zN}/\mathrm{Hz}^{1/2}$, assuming the mode mass of \SI{\sim 15}{\ng} estimated by COMSOL simulations. 

Finally, the membranes’ extremely long coherence time could enable electromechanical experiments to test fundamental physics. They may, for example, constrain the parameters of collapse models \cite{Bassi2013}, such as the continuous spontaneous localization model (CSL) \cite{Nimmrichter2014}, which is based on a nonlinear stochastic extension of the Schrödinger equation. Testing the effects of general relativity on massive quantum superpositions with such systems has also been proposed recently \cite{Gely2021}.

\section{Methods}

\label{sec:methods}
\subsection{Sample fabrication}
The planar microwave resonator is a patterned thin film of NbTiN sputter-deposited by Star Cryoelectronics on a high-resistivity silicon wafer from Topsil. The superconductor is patterned with standard UV lithography and etched with an ICP recipe based on $\mathrm{SF}_6/\mathrm{O}_2$ at low power to avoid resist burning. Aluminum pillars define the flip-chip nominal separation, and we etch a recess into the resonator Si chip using the Bosch process to minimise the risk of the flip-chip contacting anywhere else than at the pillars.

The membrane is made of stoichiometric high-stress silicon nitride patterned with standard UV lithography and etched with a $\mathrm{CF}_4/\mathrm{H}_2$-based ICP recipe on wafer front and back side. The membrane is released in a hot KOH bath. The membranes are then cleaned in a bath of piranha solution, broken off to individual chips and metallized by shadow-masked e-beam evaporation of aluminum.

\section*{Acknowledgements}

The authors would like to acknowledge support  by S.~Cherednichenko of Chalmers University in early attempts of superconductor deposition. This work was supported by the European Research Council project Q-CEOM (grant no. 638765), the Danish National Research Foundation (Center of Excellence “Hy-Q”), the EU H2020 FET proactive project HOT (grant no. 732894), as well as the Swiss National Science Foundation (grant no. 177198).  The project has furthermore received funding from the European Union’s Horizon 2020 research and innovation program under grant agreement No. 722923 (Marie Curie ETN - OMT) and the Marie Skłodowska-Curie grant agreement No 801199.

\section*{Author Contributions}

Y.~S.~designed and fabricated the microwave circuit and flip-chip assembly, performed most of the experiments and analyzed the data. T.~C.~designed the mechanical isolator, participated in  data acquisition and analysis and made the theoretical model. E.~L.~designed and fabricated the metallized phononic membrane resonator. S.~S.~and E.~P.~contributed to the experimental setup and characterisation at an early stage. Y.~S., T.~C. and A.~S. wrote the manuscript, which all authors revised. A.~S.~supervised the entire project.

\section*{Competing interests}

The authors declare no competing interests.

\section*{Data availability}

The raw data that support the findings of this study are available from the corresponding author upon reasonable request. Processed data representing all the data in the published figures, both from the main text and the supplementary material, are available in the Zenodo open repository \url{https://zenodo.org/record/5996595}, together with a Jupyter notebook to plot the figures.


\end{document}


\maketitle

\tableofcontents

\newpage

\section{Model derivation}
We are going to derive the model following a similar calculus than Weinstein \emph{et al}\cite{Weinstein2014}, but in the case of a carrier cancellation, in order to have a model for the squashing observed close to the ground state in a sideband cooling experiment.
\subsection{Langevin equations}
We start from the following Hamiltonian, which is the standard optomechanical Hamiltonian:
\begin{equation}
\hat{H}=\hbar\Omega_m\bd\bn+\hbar\omega_c\ad\an+\hbar g_0\ad\an\left(\bn+\bd\right),
\end{equation}
where $\omega_c/2\pi$ ($\Omega_m/2\pi$) is the microwave (mechanical) frequency, $g_0/2\pi$ is the microwave frequency shift induced by the zero point motion of the mechanical resonator, and $\an$ ($\bn$) is the photon (phonon) annihilation operator.
The set of Langevin equations deriving from this Hamiltonian are the following:
\begin{align}
\dt\an&=\left(-i\omega_c-\kappa/2\right)\an-ig_0\an\left(\bn+\bd\right)+\sqrt{\kappa_0}\an_\mathrm{in,0}+\sqrt{\kappa_c}\an_\mathrm{in,c}\\
\dt\bn&=\left(-i\Omega_m-\Gamma_m/2\right)\bn-ig_0\ad\an+\sqrt{\Gamma_m}\bn_\mathrm{in},
\end{align}
where $\kappa_0$ ($\kappa_c$) is the intrinsic (coupling) microwave loss rate, $\kappa\eqdef\kappa_0+\kappa_c$ is the total microwave loss rate, $\Gamma_m$ is the mechanical loss rate and $\an_\mathrm{in,0}$, $\an_\mathrm{in,c}$ and $\bn_\mathrm{in}$ are the noise bosonic operators associated with those loss channels.

We consider the case of such a system driven by a strong optical pump at the frequency $\omega_p/2\pi$, and consider the following approximations:
\begin{align}
\an&\approx e^{-i\omega_p t}\left(\alpha+\dn\right)\\
\bn&\approx \beta+\cn,
\end{align}
where $\alpha$ ($\beta$) is the amplitude of large coherent field inside the optical (mechanical) oscillator, and $\cn$ ($\dn$) is the annihilation operator corresponding to first order expansions of the field around this mean coherent state.

First, if we define $\an_r\eqdef e^{i\omega_p t}\an$, we have:
\begin{equation}
\dt\an_r=\left(i\Delta-\kappa/2\right)\an_r-ig_0\an_r\left(\bn+\bd\right)+\sqrt{\kappa_0}\an_\mathrm{in,0,r}+\sqrt{\kappa_c}\an_\mathrm{in,c,r},
\end{equation}
where $\Delta=\omega_p-\omega_c$, $\an_\mathrm{in,c,r}\eqdef\an_\mathrm{in,c}e^{i\omega_p t}$, and $\an_\mathrm{in,0,r}\eqdef\an_\mathrm{in,0}e^{i\omega_p t}$.

At zeroth order we have:
\begin{align}
0&=\left(i\Delta-\kappa/2\right)\alpha-ig_0\alpha\left(\beta+\beta^*\right)+\sqrt{\kappa_c}\alpha_\mathrm{in}\\
0&=\left(-i\Omega_m-\Gamma_m/2\right)\beta-ig_0|\alpha|^2
\end{align}
We can show that this system, provided that the incoming intensity is low enough, leads to a solution $(\alpha, \beta)$ that represents the amplitude of the mean field in the cavity.

At first order, we have:
\begin{align}
\dt\dn&=\left(i\tilde{\Delta}-\kappa/2\right)\dn-ig_0\alpha\left(\cn+\cd\right)+\sqrt{\kappa_0}\dn_\mathrm{in,0}+\sqrt{\kappa_c}\dn_\mathrm{in,c}\\
\dt\cn&=\left(-i\Omega_m-\Gamma_m/2\right)\cn-ig_0\left(\alpha\dd+\alpha^*\dn\right)+\sqrt{\Gamma_m}\cn_\mathrm{in},
\end{align}
where we have defined $\tilde{\Delta}\eqdef\Delta-g_0\left(\beta+\beta^*\right)$. We will now absorb the phase of $\alpha\eqdef|\alpha|e^{i\Psi_\alpha}$ in a redefinition of the optical annihilation operator: $\tilde{\dn}\eqdef e^{-i\Psi_\alpha}\dn$. We then have:
\begin{align}
\dt\dn&=\left(i\Delta-\kappa/2\right)\dn-ig\left(\cn+\cd\right)+\sqrt{\kappa_0}\dn_\mathrm{in,0}+\sqrt{\kappa_c}\dn_\mathrm{in,c}\\
\dt\cn&=\left(-i\Omega_m-\Gamma_m/2\right)\cn-ig\left(\dd+\dn\right)+\sqrt{\Gamma_m}\cn_\mathrm{in},
\end{align}
were the tildes were omitted for clarity, and we defined $g\eqdef g_0|\alpha|$.
\subsection{Input noise definition}
The input noise terms are defined as such:
\begin{equation}
\begin{split}
\langle\dn_\mathrm{in,0}(t)\dd_\mathrm{in,0}(t')\rangle&=\left(n_0+1\right)\delta(t-t')\\
\langle\dd_\mathrm{in,0}(t)\dn_\mathrm{in,0}(t')\rangle&=n_0\delta(t-t')\\
\langle\dn_\mathrm{in,c}(t)\dd_\mathrm{in,c}(t')\rangle&=\left(n_c+1\right)\delta(t-t')\\
\langle\dd_\mathrm{in,c}(t)\dn_\mathrm{in,c}(t')\rangle&=n_c\delta(t-t')\\
\langle\cd_\mathrm{in}(t)\cn_\mathrm{in}(t')\rangle&=\bar{n}_\mathrm{th}\delta(t-t')\\
\langle\cn_\mathrm{in}(t)\cd_\mathrm{in}(t')\rangle&=\left(\bar{n}_\mathrm{th}+1\right)\delta(t-t')\\
\end{split}
\end{equation}.
Here, $n_c$ will represent the phase noise of the system.
\subsection{Mechanical state}
We place ourselves in the so called \emph{weak coupling regime}, where the power is weak enough to have an optical loss rate well larger than the mechanical loss rate. In this condition, we can compute the mechanical occupation by assuming the optical field in its steady state, which is oscillating at the mechanical frequency. This is done by defining:
\begin{equation}
\dn\approx\dn_+e^{i\Omega_m t}+\dn_-e^{-i\Omega_m t},
\end{equation}
and that we can assume that $\dn_\pm\approx\dn e^{\mp i\Omega_m t}$. We then have:
\begin{align}
\dt\dn_+&=\left(i\left(\Delta-\Omega_m\right)-\kappa/2\right)\dn_+-ig\left(\cn+\cd\right)e^{-i\Omega_m t}+\sqrt{\kappa_0}\dn_\mathrm{in,0,+}+\sqrt{\kappa_c}\dn_\mathrm{in,c,+}\\
\dt\dn_-&=\left(i\left(\Delta+\Omega_m\right)-\kappa/2\right)\dn_--ig\left(\cn+\cd\right)e^{i\Omega_m t}+\sqrt{\kappa_0}\dn_\mathrm{in,0,-}+\sqrt{\kappa_c}\dn_\mathrm{in,c,-}.
\end{align}

In the steady state, we have $\dt\dn_\pm\sim 0$. We also have $\left(\cn+\cd\right)e^{i\Omega_m t}\sim~\cn e^{i\Omega_m t}$ and $\left(\cn+\cd\right)e^{-i\Omega_m t}\sim \cd e^{-i\Omega_m t}$. This allows to retrieve the values:
\begin{align}
\dn_+&=\mathcal{A}_-\left[-ig\cd +\hat{\mathcal{N}}_\mathrm{in}\right]e^{-i\Omega_m t}\\
\dn_-&=\mathcal{A}_+\left[-ig\cn +\hat{\mathcal{N}}_\mathrm{in}\right]e^{i\Omega_m t},
\end{align}
where we have defined $\mathcal{A}_\pm\eqdef\frac{1}{\kappa/2-i\left(\Delta\pm\Omega_m\right)}$, and $\hat{\mathcal{N}}_\mathrm{in}\eqdef\sqrt{\kappa_0}\dn_\mathrm{in,0}+\sqrt{\kappa_c}\dn_\mathrm{in,c}$.
We can inject this equation in the mechanical Langevin equation. In the limit of a high quality factor, where $\cn$ is oscillating mainly at $\Omega_m$, we have:

\begin{equation}
\dt\cn=\left(-i\Omega_m-\Gamma_m/2\right)\cn+g^2\left[\mathcal{A}_-^*-\mathcal{A}_+\right]\cn-ig\left[\mathcal{A}_+\hat{\mathcal{N}}_\mathrm{in}+\mathcal{A}_-^*\hat{\mathcal{N}}^\dagger_\mathrm{in}\right]+\sqrt{\Gamma_m}\cn_\mathrm{in}.
\end{equation}
We can rewrite the above equation in:
\begin{equation}
\dt\cn=\left(-i\Omega_\mathrm{eff}-\Gamma_\mathrm{eff}/2\right)\cn-ig\left[\mathcal{A}_+\hat{\mathcal{N}}_\mathrm{in}+\mathcal{A}_-^*\hat{\mathcal{N}}^\dagger_\mathrm{in}\right]+\sqrt{\Gamma_m}\cn_\mathrm{in},
\end{equation}
where we have:
\begin{align}
\Omega_\mathrm{eff}&\eqdef\Omega_m+\Omega_e\\
\Gamma_\mathrm{eff}&\eqdef\Gamma_m+\Gamma_e\\
\Omega_e&\eqdef -g^2\mathrm{Im}\left[\mathcal{A}_-^*-\mathcal{A}_+\right]\\
\Gamma_e&\eqdef -2g^2\mathrm{Re}\left[\mathcal{A}_-^*-\mathcal{A}_+\right].
\end{align}
This can be integrated in:
\begin{equation}
\cn(t)=\int_{-\infty}^t dt' e^{\left(i\Omega_\mathrm{eff}+\Gamma_\mathrm{eff}/2\right)\left(t'-t\right)}\left\{-ig\left[\mathcal{A}_+\hat{\mathcal{N}}_\mathrm{in}+\mathcal{A}_-^*\hat{\mathcal{N}}^\dagger_\mathrm{in}\right]+\sqrt{\Gamma_m}\cn_\mathrm{in}\right\}(t')
\end{equation}
\subsection{Optical output}
If we define the operator $\hat{X}\eqdef\left(\cn+\cd\right)$, we have for the optical field:
\begin{equation}
\dt\dn=\left(i\Delta-\kappa/2\right)\dn-ig\hat{X}+\hat{\mathcal{N}}_\mathrm{in},
\end{equation}
which can be integrated in:
\begin{equation}
\dn=\int_{-\infty}^tdt'e^{\left(-i\Delta+\kappa/2\right)\left(t'-t\right)}\left(-ig\hat{X}+\hat{\mathcal{N}}_\mathrm{in}\right)(t').
\end{equation}
Using the input-output relation:
\begin{equation}
\dn_\mathrm{out}=-\dn_\mathrm{c,in}+\sqrt{\kappa_c}\dn,
\end{equation}
which, after carrier cancellation is changed into:
\begin{equation}
\dn_\mathrm{out}=+\sqrt{\kappa_c}\dn,
\end{equation}
and assuming that we measure the operator $\hat{I}\eqdef\dn_\mathrm{out}+\dd_\mathrm{out}$, we have:
\begin{align}
\hat{I}(t)=&+\sqrt{\kappa_c}\int_{-\infty}^tdt'e^{\kappa/2\left(t'-t\right)}\Bigg(g\hat{X}(t')\left(-ie^{-i\Delta\left(t'-t\right)}+ie^{i\Delta\left(t'-t\right)}\right)\\
&+e^{-i\Delta\left(t'-t\right)}\hat{\mathcal{N}}_\mathrm{in}(t')\\
&+e^{i\Delta\left(t'-t\right)}\hat{\mathcal{N}}_\mathrm{in}^\dagger(t')\Bigg)
\end{align}
\subsection{Spectrum output}
We will use the same spectrum definition as in \cite{Weinstein2014}, namely:
\begin{equation}
\mathcal{S}[\omega]\eqdef\frac{1}{2}\int_{-\infty}^\infty dt\langle I(t)I(0)+I(0)I(t)\rangle e^{-i\omega t}
\end{equation}
We first compute the measured power:
\begin{equation}
\frac{1}{2}\langle \{I(t),I(0)\}\rangle=\mathcal{P}_\mathrm{shot, shot}+\mathcal{P}_\mathrm{mech, mech}+\mathcal{P}_\mathrm{mech, shot},
\end{equation}
where $\{A,B\}\eqdef AB+BA$, $\mathcal{P}_\mathrm{shot, shot}$ is the contribution from the optical noise, $\mathcal{P}_\mathrm{mech, mech}$ is the contribution from the mechanical noise and $\mathcal{P}_\mathrm{mech, shot}$ is the cross spectrum.
\subsubsection{Shot noise}
We have:
\begin{equation}
\begin{split}
\mathcal{P}_\mathrm{shot, shot}=\frac{1}{2}\langle &\{\sqrt{\kappa_c}\int_{-\infty}^tdt'e^{\kappa/2\left(t'-t\right)}\Bigg[e^{-i\Delta\left(t'-t\right)}\hat{\mathcal{N}}_\mathrm{in}(t')\\
&+e^{i\Delta\left(t'-t\right)}\hat{\mathcal{N}}_\mathrm{in}^\dagger(t')\Bigg],\\
&+\sqrt{\kappa_c}\int_{-\infty}^0dt'e^{\kappa/2t'}\Bigg[e^{-i\Delta t'}\hat{\mathcal{N}}_\mathrm{in}(t')\\
&+e^{i\Delta t'}\hat{\mathcal{N}}_\mathrm{in}^\dagger(t')\Bigg]\}\rangle,
\end{split}
\end{equation}
which reads:
\begin{equation}
\mathcal{P}_\mathrm{shot, shot}=\kappa \eta\left(\tilde{n}+1/2\right)e^{-\kappa/2|t|}\left[e^{i\Delta t}+e^{-i\Delta t}\right],
\end{equation}
where $\tilde{n}=\left(1-\eta\right)n_0+\eta n_c$. This leads to the following noise floor:
\begin{equation}
\begin{split}
\mathcal{S}[\omega]_\mathrm{shot, shot}&=\eta\kappa^2\left(\tilde{n}+1/2\right)\Bigg[\frac{1}{\left(\kappa/2\right)^2+\left(\omega-\Delta\right)^2}+\frac{1}{\left(\kappa/2\right)^2+\left(\omega+\Delta\right)^2}\Bigg]
\end{split}
\end{equation}

\subsubsection{Mechanical noise}
We have, in the limit of a high quality factor:
\begin{equation}
\langle \hat{X}(t), \hat{X}(t')\rangle\approx\langle \cn(t), \cd(t')\rangle+\langle \cd(t), \cn(t')\rangle
\end{equation}
This gives:
\begin{equation}
\begin{split}
\langle \hat{X}(t), \hat{X}(t')\rangle&\approx\frac{\Gamma_m}{\Gamma_\mathrm{eff}}e^{-\Gamma_\mathrm{eff}/2|t'-t|}\left[\left(\bar{n}_\mathrm{th}+1\right) e^{i\Omega_\mathrm{eff}\left(t'-t\right)}+\bar{n}_\mathrm{th}e^{-i\Omega_\mathrm{eff}\left(t'-t\right)}\right]\\
&+e^{-\Gamma_\mathrm{eff}/2|t-t'|}e^{-i\Omega_\mathrm{eff}\left(t-t'\right)}\frac{g^2\kappa}{\Gamma_\mathrm{eff}}\Bigg\{|\mathcal{A}_+|^2\left(\tilde{n}+1\right)+|\mathcal{A}_-|^2\tilde{n}\Bigg\}\\
&+e^{-\Gamma_\mathrm{eff}/2|t-t'|}e^{i\Omega_\mathrm{eff}\left(t-t'\right)}\frac{g^2\kappa}{\Gamma_\mathrm{eff}}\Bigg\{|\mathcal{A}_-|^2\left(\tilde{n}+1\right)+|\mathcal{A}_+|^2\tilde{n}\Bigg\}
\end{split}
\end{equation}
This expression allows to compute the part of the spectrum associated with the mechanical noise. We have:
\begin{equation}
\begin{split}
\mathcal{P}_\mathrm{mech, mech} (t)&=\frac{g^2\kappa_c}{\Gamma_\mathrm{eff}}e^{-\Gamma_\mathrm{eff}|t|}\left(e^{i\Omega_\mathrm{eff}t}+e^{-i\Omega_\mathrm{eff} t}\right)\left(|\mathcal{A}_+|^2+|\mathcal{A}_-|^2\right)\\
&\times\left[\Gamma_m\left(n_m+1/2\right)+g^2\kappa\left(|\mathcal{A}_+|^2+|\mathcal{A}_-|^2\right)\left(\tilde{n}+1/2\right)\right]
\end{split},
\end{equation}
which in turn gives the following spectrum:
\begin{equation}
\begin{split}
\mathcal{S}[\omega]_\mathrm{mech, mech}&=g^2\kappa_c\left(\frac{1}{\left(\Gamma_\mathrm{eff}/2\right)^2+\left(\omega-\Omega_\mathrm{eff}\right)^2}+\frac{1}{\left(\Gamma_\mathrm{eff}/2\right)^2+\left(\omega+\Omega_\mathrm{eff}\right)^2}\right)\left(|\mathcal{A}_+|^2+|\mathcal{A}_-|^2\right)\\
&\times\left[\Gamma_m\left(\bar{n}_\mathrm{th}+1/2\right)+g^2\kappa\left(|\mathcal{A}_+|^2+|\mathcal{A}_-|^2\right)\left(\tilde{n}+1/2\right)\right]
\end{split}
\end{equation}
\subsubsection{Cross spectrum}
The last part of the spectrum is the cross spectrum between the input optical noise and the backaction noise.
We have:
\begin{equation}
\langle I(t), I(0)\rangle_\mathrm{cross}=\langle I(t), I(0)\rangle_\mathrm{cross, 1}+\langle I(t), I(0)\rangle_\mathrm{cross, 2},
\end{equation}
where:
\small{
\begin{equation}
\begin{split}
\langle I(t), I(0)\rangle_\mathrm{cross, 1}&\eqdef\langle \sqrt{\kappa_c}\int_{-\infty}^tdt'e^{\kappa/2\left(t'-t\right)}\left[e^{-i\Delta\left(t'-t\right)}\hat{\mathcal{N}}_\mathrm{in}(t')+e^{i\Delta\left(t'-t\right)}\hat{\mathcal{N}}_\mathrm{in}^\dagger(t')\right],\\
&g\sqrt{\kappa_c}\int_{-\infty}^0dt''e^{\kappa/2t''}\left(-ie^{-i\Delta t''}+ie^{i\Delta t''}\right)\\
&\Bigg\{-ig\int_{-\infty}^{t''}dt'''e^{\left(\Gamma_\mathrm{eff}/2+i\Omega_\mathrm{eff}\right)\left(t'''-t''\right)}\left(\mathcal{A}_+\hat{\mathcal{N}}_\mathrm{in}+\mathcal{A}_-^*\hat{\mathcal{N}}_\mathrm{in}^\dagger\right)(t''')\\
&+ig\int_{-\infty}^{t''}dt'''e^{\left(\Gamma_\mathrm{eff}/2-i\Omega_\mathrm{eff}\right)\left(t'''-t''\right)}\left(\mathcal{A}_+^*\hat{\mathcal{N}}_\mathrm{in}^\dagger+\mathcal{A}_-\hat{\mathcal{N}}_\mathrm{in}\right)(t''')\Bigg\}\rangle,
\end{split}
\end{equation}}

and:
\begin{equation}
\begin{split}
\langle I(t), I(0)\rangle_\mathrm{cross, 2}&\eqdef\langle g\sqrt{\kappa_c}\int_{-\infty}^tdt'e^{\kappa/2\left(t'-t\right)}\left(-ie^{-i\Delta \left(t'-t\right)}+ie^{i\Delta \left(t'-t\right)}\right)\\
&\Bigg\{-ig\int_{-\infty}^{t'}dt''e^{\left(\Gamma_\mathrm{eff}/2+i\Omega_\mathrm{eff}\right)\left(t''-t'\right)}\left(\mathcal{A}_+\hat{\mathcal{N}}_\mathrm{in}+\mathcal{A}_-^*\hat{\mathcal{N}}_\mathrm{in}^\dagger\right)(t'')\\
&+ig\int_{-\infty}^{t'}dt''e^{\left(\Gamma_\mathrm{eff}/2-i\Omega_\mathrm{eff}\right)\left(t''-t'\right)}\left(\mathcal{A}_+^*\hat{\mathcal{N}}_\mathrm{in}^\dagger+\mathcal{A}_-\hat{\mathcal{N}}_\mathrm{in}\right)(t'')\Bigg\},\\
&\sqrt{\kappa_c}\int_{-\infty}^0dt'''e^{\kappa/2t'''}\left[e^{-i\Delta t'''}\hat{\mathcal{N}}_\mathrm{in}(t''')+e^{i\Delta t'''}\hat{\mathcal{N}}_\mathrm{in}^\dagger(t''')\right]\rangle,
\end{split}
\end{equation}
with similar definitions for $\langle I(0), I(t)\rangle$. The computation gives, in the resolved sideband and the weak coupling regime:
\begin{equation}
\begin{split}
\mathcal{P}_\mathrm{mech, shot} (t)=g^2\kappa^2\eta e^{-\Gamma_\mathrm{eff}|t|/2}&\left(e^{i\Omega_\mathrm{eff}t}+e^{-i\Omega_\mathrm{eff}t}\right)\left(\mathcal{A}_--\mathcal{A}_+^*\right)\\
&\times\left(|\mathcal{A}_-|^2+|\mathcal{A}_+|^2\right)\left(\tilde{n}+1/2\right)
\end{split}
\end{equation}
which in turn gives the following spectrum:
\begin{equation}
\begin{split}
\mathcal{S}[\omega]_\mathrm{mech, shot}&=g^2\kappa^2\eta\Bigg\{\left(\frac{1}{\left(\Gamma_\mathrm{eff}/2\right)^2+\left(\omega-\Omega_\mathrm{eff}\right)^2}+\frac{1}{\left(\Gamma_\mathrm{eff}/2\right)^2+\left(\omega+\Omega_\mathrm{eff}\right)^2}\right)\\
&\times\Gamma_\mathrm{eff}\left(\mathcal{A}_--\mathcal{A}_+^*\right)\big[\left(|\mathcal{A}_-|^2+|\mathcal{A}_+|^2\right)\left(\tilde{n}+1/2\right)\big]
\end{split}
\end{equation}
\subsection{Final result}
In the case of a red detuned beam, in the resolved sideband regime, we have $\Delta\approx -\Omega_m$, $|\mathcal{A}_-|^2\approx 0$, $|\mathcal{A}_+|^2\approx 4/\kappa^2$ and therefore:
\begin{equation}
\begin{split}
S[\omega]=n_\mathrm{add}+4\eta\left(\tilde{n}+1/2\right)+\eta \Gamma_\mathrm{m}\Gamma_\mathrm{e}\frac{\bar{n}_\mathrm{th}+\frac{1}{2}-\left(2+\frac{\Gamma_\mathrm{e}}{\Gamma_\mathrm{m}}\right)\left(\tilde{n}+\frac{1}{2}\right)}{\left(\Gamma_\mathrm{m}+\Gamma_\mathrm{e}\right)^2/4+\left(\omega-\omega_\mathrm{p}-\Omega_\mathrm{eff}\right)^2}
\end{split}
\end{equation}.

\section{System Summary}

\begin{table}[h!]
    \centering
\begin{tabular}{c|c}
    Microwave cavity frequency & $\omega_c / 2\pi = \SI{8.350}{\GHz} $ \\
    Cavity total decay rate &  $\kappa / 2\pi = \SI{226}{\kHz} $ \\
    Cavity out-coupling rate &  $\kappa_e / 2\pi = \SI{183}{\kHz} $ \\
    Cavity internal loss rate &  $\kappa_i / 2\pi = \SI{43}{\kHz} $ \\
    Cavity coupling efficiency &  $\eta_c =  0.81$ \\
    
    Mechanical frequency &  $\Omega_m / 2\pi = \SI{1.487}{\MHz} $ \\
    Mechanical energy decay rate &  $\Gamma_m / 2\pi = \SI{1.0}{\mHz} $ \\
    Mechanical quality factor &  $Q_m = \Omega_m / \Gamma_m = 1.5\cdot 10^9$ \\
    
    Electro-mechanical single photon-coupling rate & $g_0 = \SI{0.89 \pm 0.11}{\Hz} $
\end{tabular}
    \caption{Experimentally measured electro-mechanical system parameters}
    \label{tab:params}
\end{table}

\begin{figure}
    \centering
    \includegraphics[width=1.\linewidth]{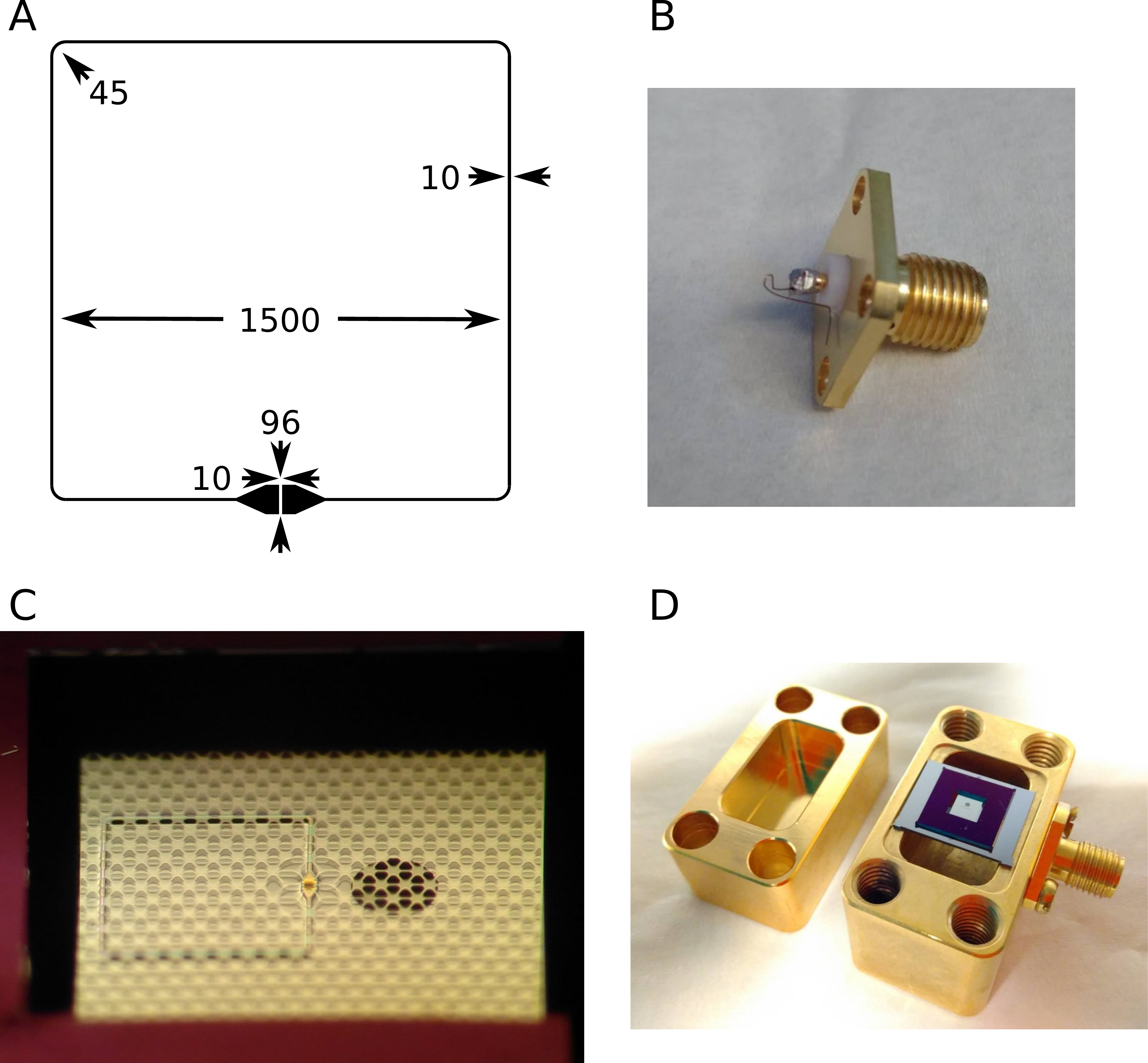}

\caption{A) Loop-gap resonator mask with dimension in microns. B) SMA connector for inductive in-coupling to the microwave cavity: one end of the hand-wound loop is soldered to the central pin of the connector, while the other end is clamped on the sample holder for grounding. C) Zoom-in at an angle of the assembled flipchip with the patterned membrane hovering over the loop-gap resonator. D) The flipchip is placed in the middle of a hollow volume of a copper (gold-coated) sample holder. The inductive in-coupling loop (shown in B) protrudes into the hollow volume underneath the resonator to achieve the mutual inductive coupling.}
    \label{fig:smaconn}
\end{figure}

The microwave device consists of a superconducting loop, whose dimensions are sketched in Fig.~\ref{fig:smaconn}A. The loop features a gap on its bottom side which is bridged by a superconducting pad deposited on a membrane (see Fig.1A of the main text). The vertical distance of the pad to the loop electrodes is $d$: the exact value of $d$ will determine the microwave resonance frequency and the coupling of microwaves and mechanics. The device is placed in the vacuum volume of a metal box. The read-out of the microwave resonator is done by inductive coupling to a loop antenna (see Fig.~\ref{fig:smaconn}B) which terminates a coaxial cable (grounded at the box) leading to our instrumentation.

\subsection{Finite element simulations}

\begin{figure}
    \centering
    \includegraphics[width=1.\linewidth]{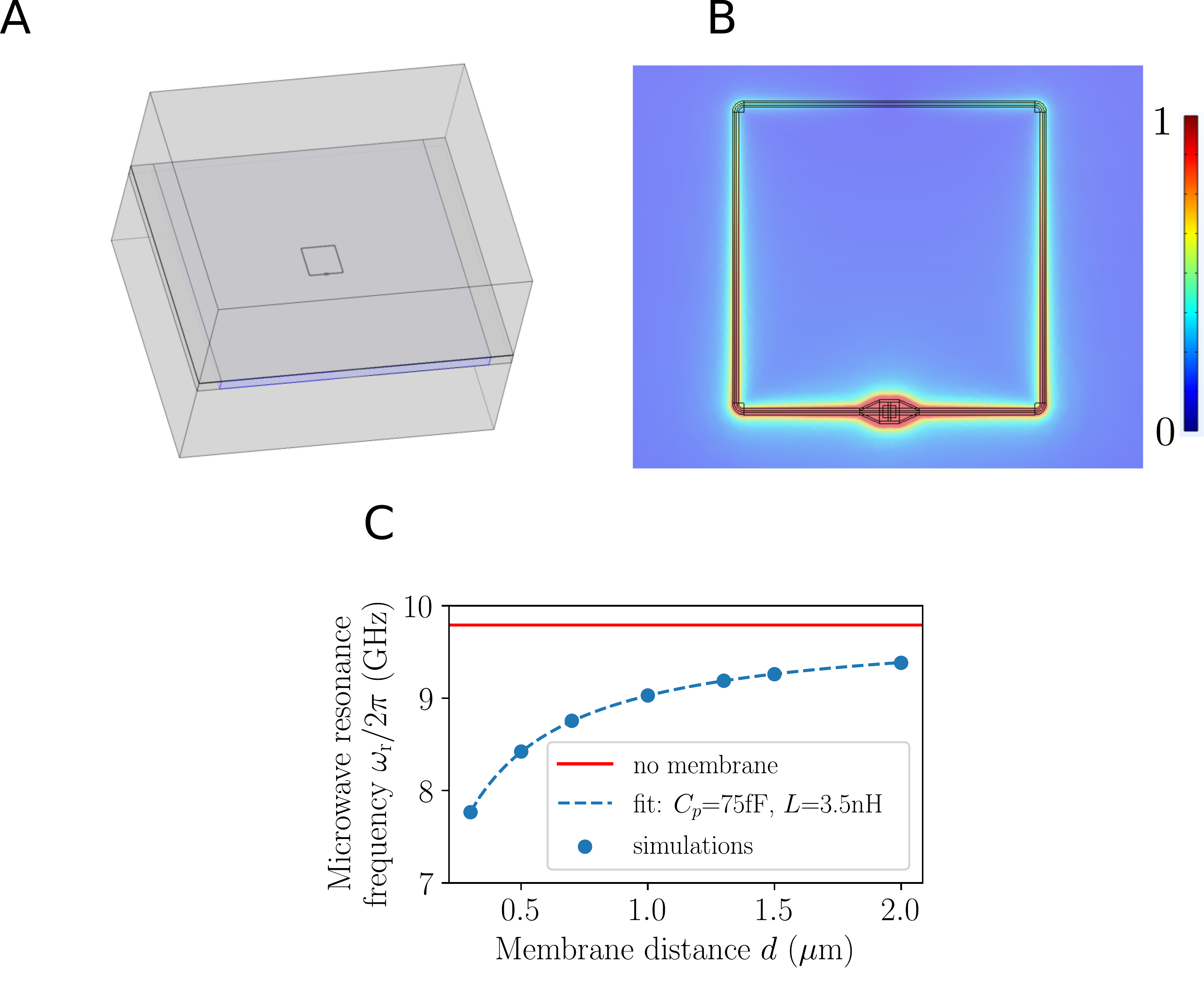}

\caption{A) Finite element geometry to simulate the microwave mode. B) Simulated electric field norm for the loop fundamental mode. C) As the membrane distance is reduced, the microwave resonance frequency is pulled-down. Simulated frequencies are fitted with Eq.~(\ref{eq:omegar}).}
    \label{fig:sims}
\end{figure}

In the commercial finite element solver COMSOL, we simulate the microwave device to estimate the resonance frequency. The geometry is depicted Fig.~\ref{fig:sims}A: the loop-gap resonator is on a silicon substrate (highlighted in light blue). The fundamental mode of the loop (depicted in Fig.~\ref{fig:sims}B) has currents flowing back and forth around the loop with the largest electric field across the pad/electrode gap. The concentration of electric field at the location of the mechanical element allows for high electro-mechanical coupling.

In the simulations, we vary the membrane-pad-to-electrodes distance $d$ and fit its effect on the resonance frequency (Fig.~\ref{fig:sims}C). As $d$ is reduced, the parallel-plate capacitance $C_\text{m}$ formed by the pad and the electrodes increases which pulls down the microwave frequency $\omega_\text{r}$. We model the mechanical capacitance to be in parallel with the loop's own capacitance $C_\text{p}$, this total capacitance with the loop inductance forms the LC resonator with frequency
\begin{align}
    \omega_\text{r} = \frac{1}{\sqrt{[C_\text{m}(d) + C_\text{p}] L}}.
    \label{eq:omegar}
\end{align}
The fit of simulated frequencies allows us to extract the parasitic capacitance $C_\text{p}$: for the fabricated geometry, we simulate $C_\text{p} \approx \SI{75}{\femto \F}$. In absence of membrane pad, the simulated loop resonance is at approx. \SI{9.8}{\GHz}.

From the measured microwave frequency in Fig.~\ref{fig:cavfit} $\omega_\text{r} = \SI{8.349}{\GHz}$, we can estimate $d \approx \SI{450}{\nm}$. The measured participation ratio of capacitances $C_\text{m}/(C_\text{m}+C_\text{p}) \approx 0.25$.

\pagebreak

\section{Calibrations}

\subsection{Example of a cavity fit}

\begin{figure}
    \centering
    \includegraphics[width=1.\linewidth]{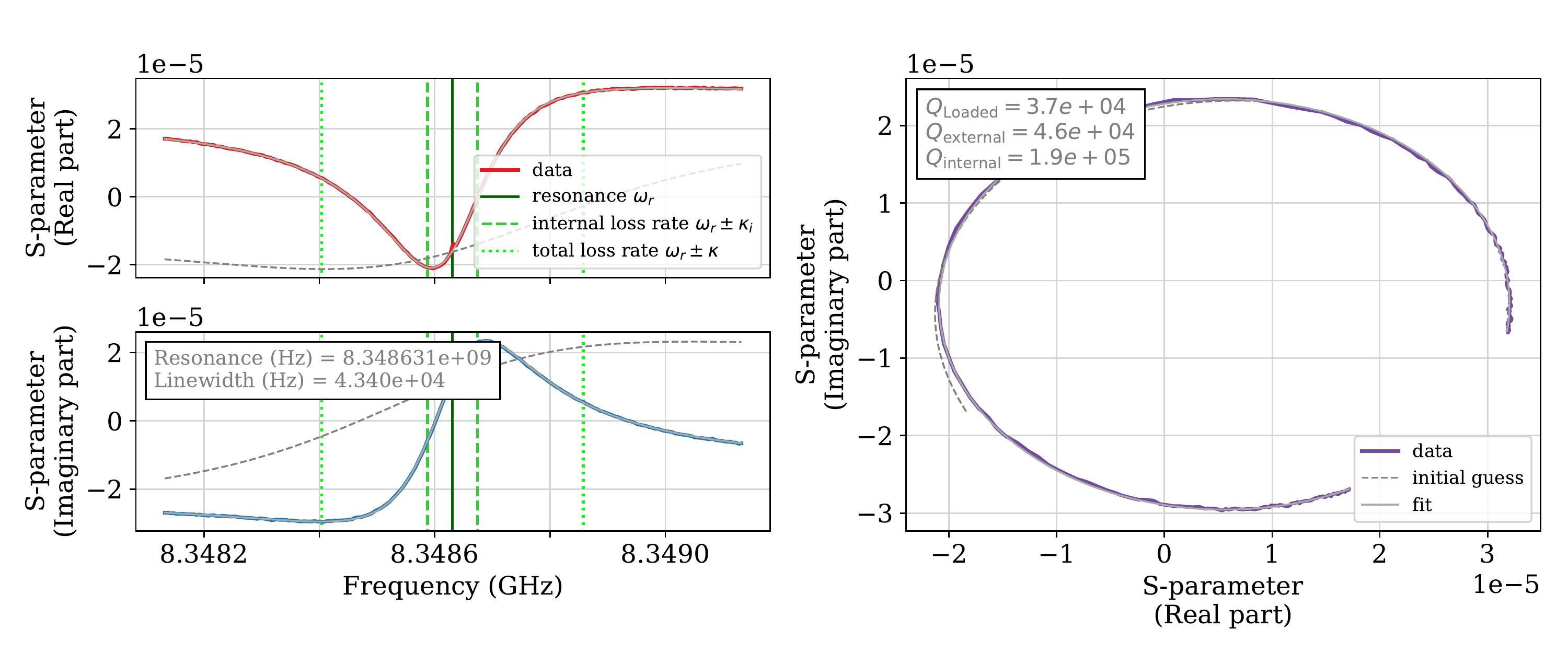}

\caption{Example of a cavity fit: here for the data point at 0 dBm from Fig.3D in the main text, where the mechanics is prepared in its ground state.}
    \label{fig:cavfit}
\end{figure}

In Fig.~\ref{fig:cavfit}, we plot the measured cavity reflection coefficient as function of frequency and fit it to extract cavity parameters. This spectroscopy allows gives us the parameters in Table~\ref{tab:params}, and is taken for the data point in Fig.~3D, where the mechanics is prepared in its ground state.

\subsection{Noise photons in the cavity}

\begin{figure}
    \centering
    \includegraphics[width=.6\linewidth]{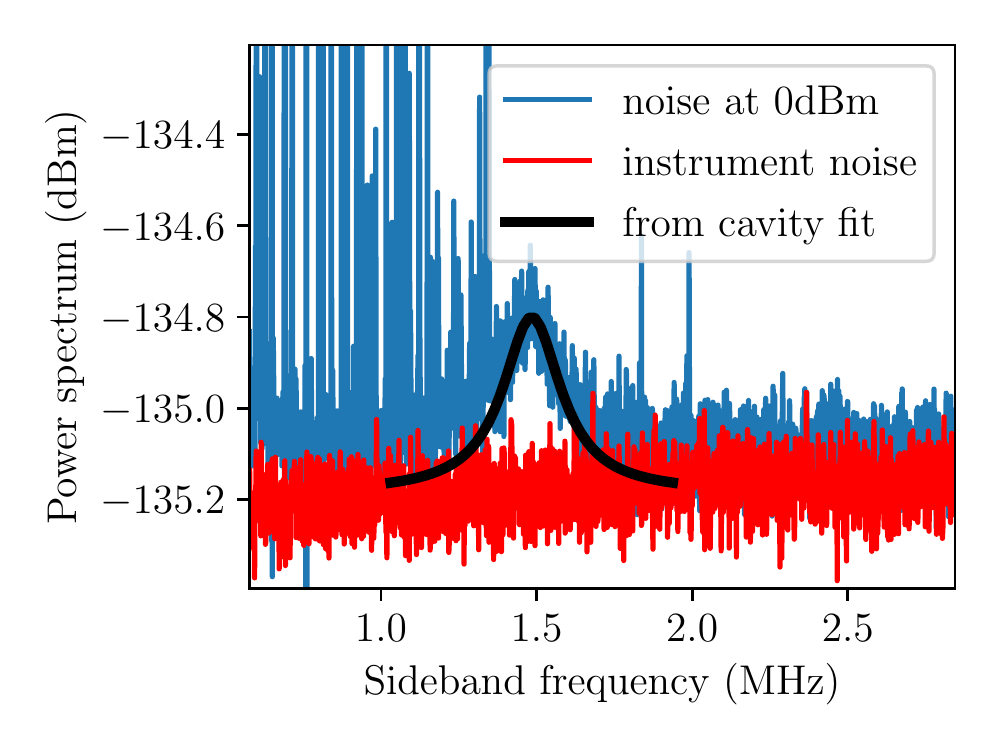}

\caption{Spectrum at the mechanical bandgap at high pump power. At high powers, a wide feature appears which has the lineshape of the cavity. This cavity feature results from cavity population, which could be caused by relative frequency fluctuations of the pump and the cavity.}
    \label{fig:cavnoise}
\end{figure}

To confirm the origin of the increased background level around the mechanical feature (as seen in Fig.~3B of the main text), which leads to noise squashing, we plot the wider power spectrum around the mechanical bandgap in Fig.~\ref{fig:cavnoise} corresponding to the ground state preparation data point in Fig.~3D of the main text. A wide feature appears at the cavity frequency which we overlay with a cavity lineshape function $\text{ls}(\omega) \propto [1 + (\omega-\omega_\text{r})^2/\kappa^2]^{-1}$ on top of the instrument background. 
In this sideband spectrum, the cavity appears in the mechanical bandgap at the location of the mechanical mode frequency because the pump is red-detuned by $\Omega_\mathrm{m}$.
In this lineshape function we use $\omega_\text{r}$ and $\kappa$ from the vector network analyser fit in Fig.~\ref{fig:cavfit}. 
As reference, we also plot a low pump power spectrum from which we extract the instrument noise background.

Following ref.~\cite{Yuan2015}, we can interpret the excess noise as a relative phase noise between the pump and the cavity, and obtain a value of this phase noise to be approx.~-145~dBc/Hz, which is about 10~dB higher than the phase noise as specified (at 1~MHz sideband frequency) by the manufacturer of our low noise signal generator. This discrepancy makes us suspect that the dominant phase noise actually stems from the cavity, such as cavity frequency jitter. Indeed we have tried adding a microwave notch filter at the exit of the signal generator, reducing its phase noise by an additional 8~dB, but this did not lead to improved cooling limit of the mechanics.

\subsection{Additional mechanical dephasing}

The high signal-to-noise of the ringdown measurements presented in Fig.2 of the main text allows us to also look at the frequency stability of the mechanical resonator. We specifically look at the phase $\phi(t)$ of the three ringdowns at powers \num{-60}, \num{-50} and \SI{-40}{dBm} and compute their instantaneous frequency $d\phi(t)/dt$ (with respect to the spectrum analyser's local oscillator) over the \SI{\approx 600}{\s} of the ringdown. We choose these three ringdowns since they are the only ones of this measurement series which are far above the noise background for the entirety of the ringdown. 

\begin{figure}
    \centering
    \includegraphics[width=1.\linewidth]{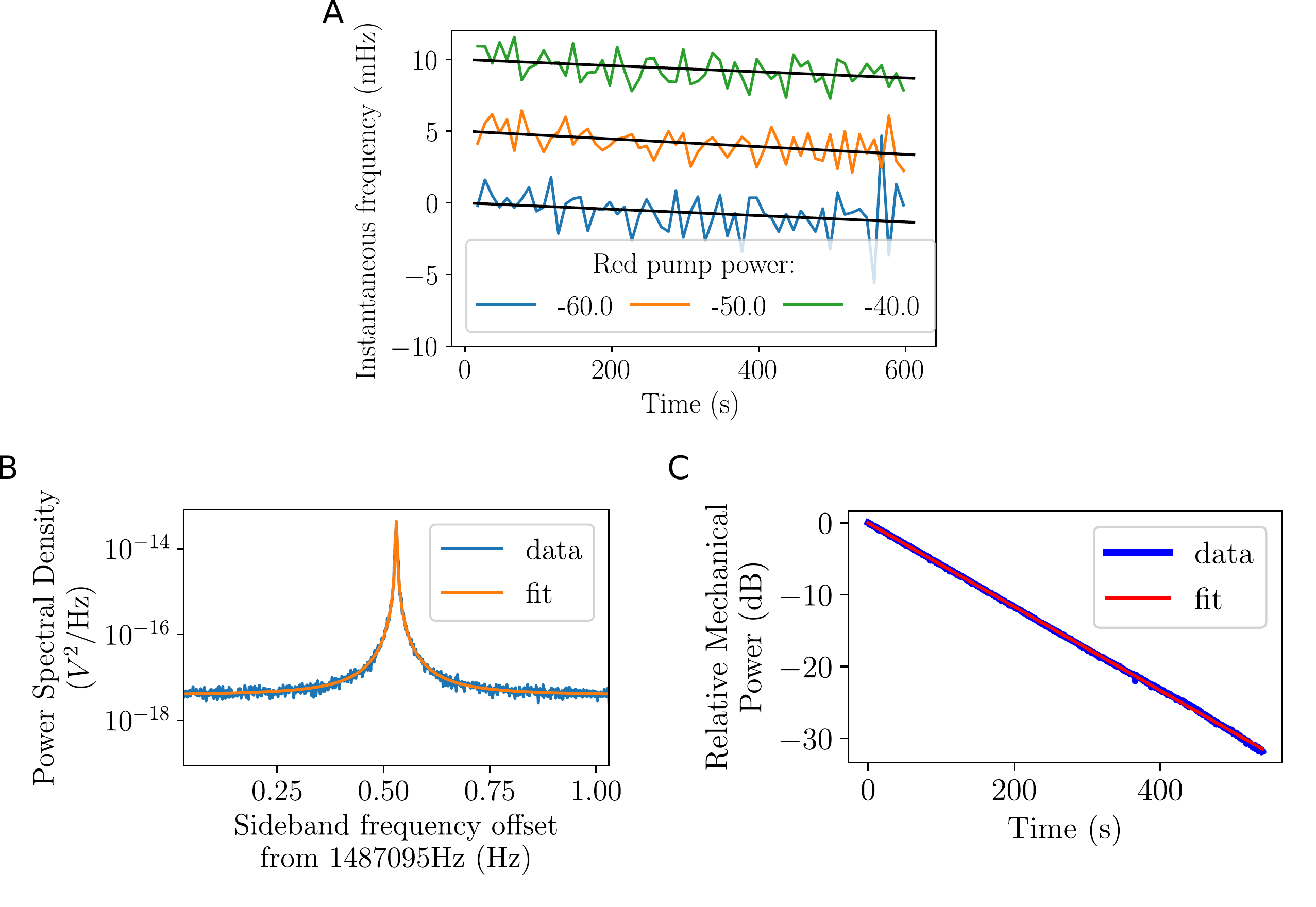}

\caption{A) Instantaneous frequency during ringown plotted over time as offset from its value $f_0$ at time at $t=0$, traces are offset by \SI{5}{\mHz} for clarity. A small drift of the mechanical frequency is visible, the drift after \SI{600}{s} is at most \SI{1.6}{\mHz}. B) Mechanical spectrum of a second device fitted with a Lorentzian function gives a linewidth of \SI{2.60 \pm 0.15}{\mHz}. C) Energy ringdown for the second device taken in the same conditions as the spectrum yields a decay rate of \SI{2.13 \pm 0.02}{\mHz}.}
    \label{fig:ringphase}
\end{figure}

In Fig.\ref{fig:ringphase}A, we plot the instantaneous mechanical frequency over time (averaged over \SI{10}{s} for each data point) of each ringdown and fit them by an affine function $f(t) = d_0 t + f_0$, with $d_0$ the drift over time and $f_0$ the initial frequency. We extract a small drift of \SI{2.7}{\micro \Hz \per \s} on all three ringdowns, corresponding to a drift \SI{1.6}{\mHz} after \SI{600}{s}. The extracted $f_0$ agree within \SI{3}{\mHz}, consistent with the drift.

We have not taken spectra with enough frequency resolution to plot the mechanical spectrum and extract spectral linewidth for the device in this paper. 
However on a different device, we have compared a large signal-to-noise ratio mechanical spectrum with an energy ringdown measurement in the same conditions. 
Spectrum and ringdown for this second device are plotted in Fig.\ref{fig:ringphase}B and C respectively. 
The linewidth from the Lorentzian spectral fit and the energy decay rate from ringdown are \SI{2.60 \pm 0.15}{\mHz} and  \SI{2.13 \pm 0.02}{\mHz}: thus any additional dephasing makes up at most $22\%$ of the total mechanical linewidth. 
This second device in question had a mechanical frequency of $\Omega_\mathrm{m}/2\pi = \SI{1.487}{\MHz}$ and therefore a spectral and ringdown quality factor of \num{ 570 \pm 33}M and \num{696 \pm 5}M.

\subsection{Gorodetsky calibration}

To calibrate the single photon coupling rate $g_0$, we use the so-called ``Gorodetsky method''\cite{Gorodetksy2010} where we compare the phase-modulation imparted by the mechanical modulation of the cavity frequency to the pump with a known reference phase-modulation (See Fig.~\ref{fig:ringphase}A). This method requires that the amount of mechanical quanta is known, which we extract by confirming that the mechanics is in a thermal state. The number of quanta in the mechanical thermal state can then be computed using the Bose-Einstein relation.

The Gorodetsky calibration is set up as an unbalanced homodyne measurement. We phase modulate the pump internally in the signal generator, split it with a 3dB splitter where half of the signal goes to the electro-mechanics cavity and the other half is used as the homodyne local oscillator. Attenuation in the dilution refrigerator is large such that the LO amplitude is much higher than the signal returned from the microwave cavity. We additionally use a variable phase shifter on the LO to optimise the interference of cavity signal with LO. A wiring diagram is shown in Fig.~\ref{fig:setup}A. The homodyne measurement is saved as the I and Q quadratures of the electric field, for which we numerically take the modulus squared to plot the power spectrum, as in Fig.~\ref{fig:ringphase}A.

We apply a pump on cavity resonance, red-detuned by a few kHz, such that, if there is any dynamical backaction, the mechanics does not become unstable and instead is slightly broadened. We monitor the amount of backaction by measuring a ringdown at the pump power of the spectral measurement and compare it to a ringdown at low power, where we know the mechanics to not be broadened (See Fig.\ref{fig:ringphase}B).

In Fig.\ref{fig:ringphase}C we plot the ratio of mechanical peak area to calibration peak area, extracted from spectra such as in Fig.\ref{fig:ringphase}A. At each temperature point this ratio is multiplied by $(\Gamma_\text{opt} + \Gamma_\text{m}) / \Gamma_\text{m}$ corresponding to the cooling factor due to dynamical backaction. The linear relationship of measurement points to thermometer reading $T$ at high temperatures indicates the mechanics is thermalised at least above \SI{200}{\milli \K}. We can use this thermalisation condition of extract the number of mechanical quanta with $\bar{n} \approx k_\text{B} T / \hbar \Omega_\text{m}$, with $k_\text{B}$ Boltzmann's constant and $\hbar$ Planck's constant divided by $2\pi$.
Below \SI{200}{\milli \K}, the mechanical area has a large scatter which the authors in Ref.~\cite{Zhou2019a} attribute to a yet unknown force which appear at very low temperatures.

The calibration yields $g_0 = \SI{0.89 \pm 0.11}{\Hz}$.

\begin{figure}
    \centering
    \includegraphics[width=1.\linewidth]{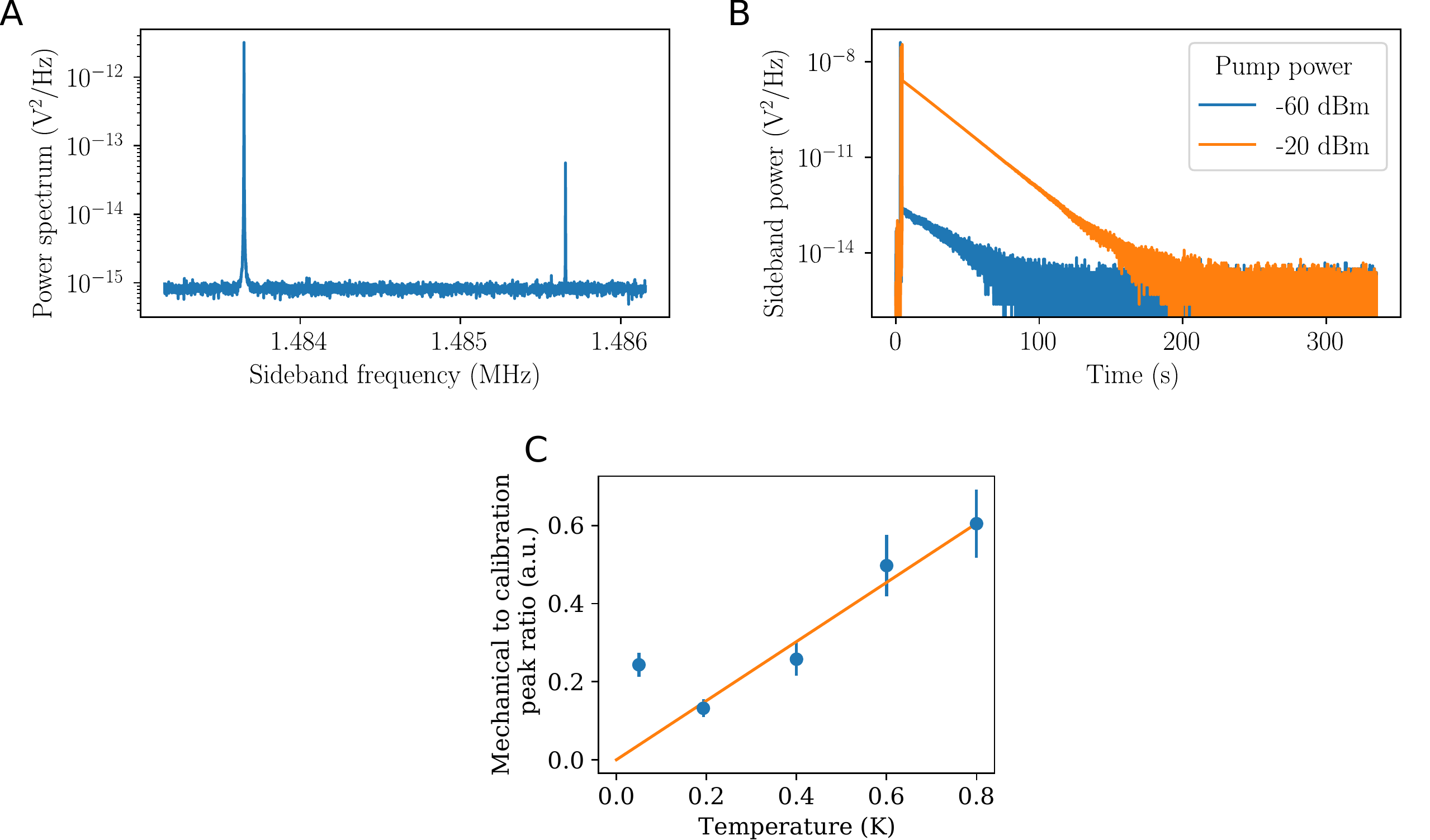}

\caption{A) An example spectrum with phase-modulated calibration peak (lower frequency) and mechanical thermal peak (upper frequency) at \SI{800}{\milli \K} and -20 dB drive B) Mechanical ringdown measurements without dynamical backaction (blue) and at the pump power of the spectrum in A (orange), where there is a small amount of backaction. C) Ratio of mechanical peak area to calibration peak area for varying sample temperature, where we adjust for any dynamical backaction using the ringdown measurements, such as in B. The line is a linear fit to the higher temperature points, confirming thermalisation of the mechanics to the environment above \SI{200}{\milli \K}. Error bars correspond to one std. dev. of the Lorentzian fit.}
    \label{fig:ringphase}
\end{figure}


\subsection{Cavity monitoring}

\begin{figure}[h!]
    \centering
    \includegraphics[width=1.\linewidth]{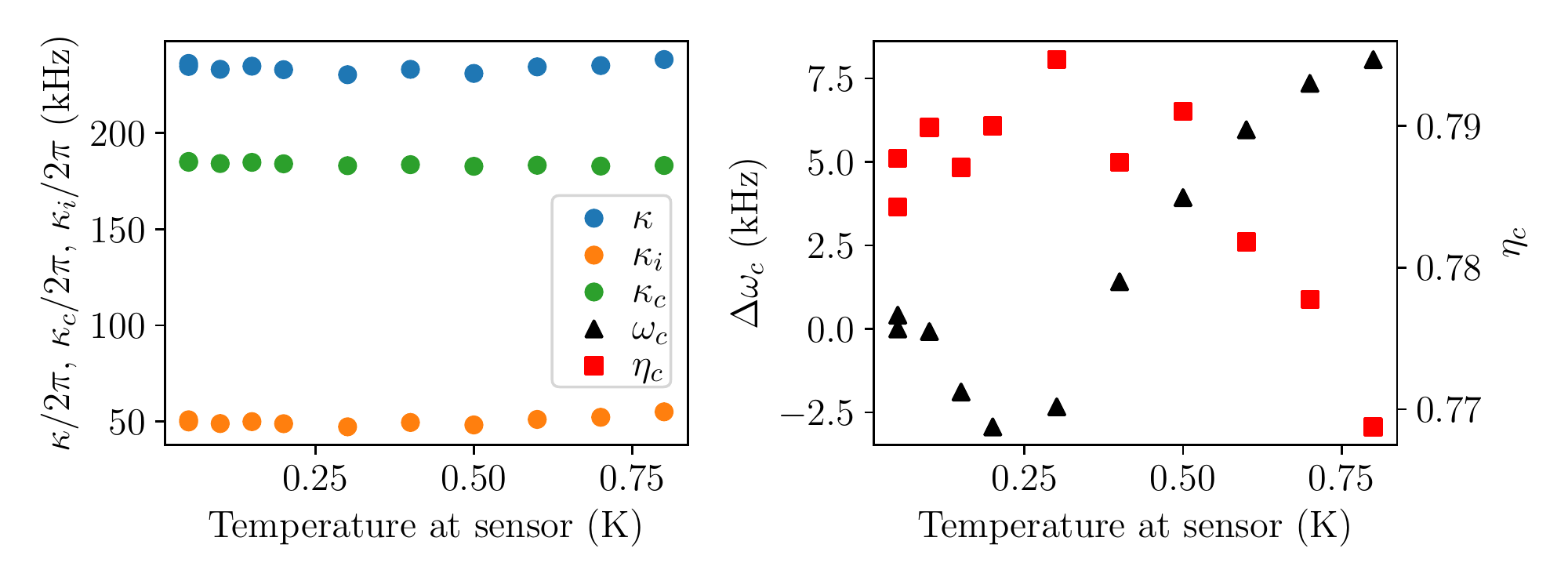}

\caption{Changes in cavity parameters as the sample temperature is varied, during the calibration sequence. Cavity scans are taken with \SI{-45}{\dBm} output from the signal generator.}
    \label{fig:cavtemp}
\end{figure}

\begin{figure}[h!]
    \centering
    \includegraphics[width=1.\linewidth]{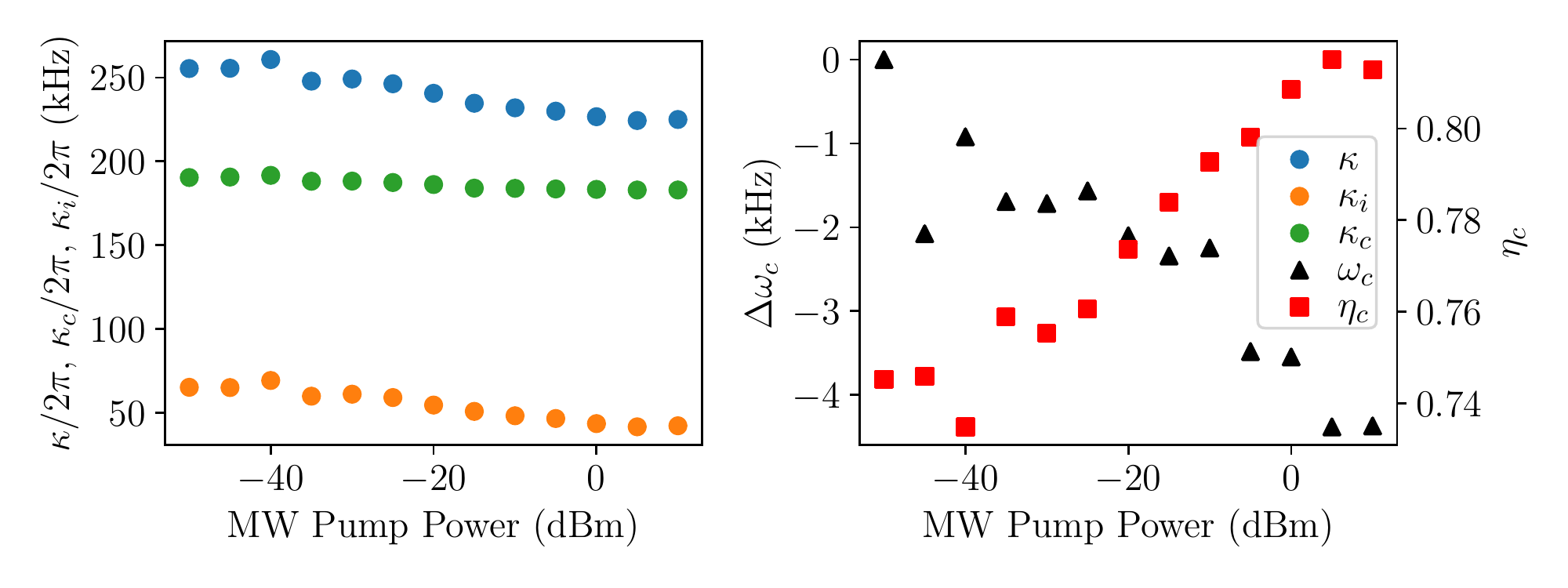}

\caption{Changes in cavity parameters as the pump power is increased, during the ground state cooling series. All cavity measurements are taken at \SI{30}{\milli \K}.}
    \label{fig:cavpower}
\end{figure}

In the measurement series for thermal calibration in Fig.~2A of the main text as well as for the ground state cooling in Fig.~3, we monitor the cavity parameters for each temperature and power point. 
Cavity parameters are plotted as function of temperature in Fig.~\ref{fig:cavtemp} and as function of power at base temperature in Fig.~\ref{fig:cavpower}.
We then use these updated parameters to extract the mechanical occupation at each temperature and power.

At higher pump powers in Fig.~\ref{fig:cavpower}, we see a reduction of the internal loss and we see a logarithmic dependence of the resonance frequency to power. Both of effect suggest the presence of two-level systems in the surrounding material coupling to the microwave field\cite{Capelle2020}. Furthermore these measurements confirm the cavity remains linear at all pump powers: that is the kinetic inductance of the NbTiN thin film is too small to impact the resonance frequency. If there were a kinetic inductance non-linearity, the resonance frequency would go linearly with power and not logarithmically.

Additionally, we do not see a significant increase in internal cavity loss as the sample temperature moves towards the superconducting transition temperature (around \SI{1}{\K}) of the Al membrane metallisation. A low participation ratio of mechanically-mediated capacitance to total capacitance could explain why the microwave losses are not affected by the superconducting state of the aluminum.

\subsection{Mechanical mode damping}

\begin{figure}[h!]
    \centering
    \includegraphics[trim=0 0 0 0, clip, width=.7\linewidth]{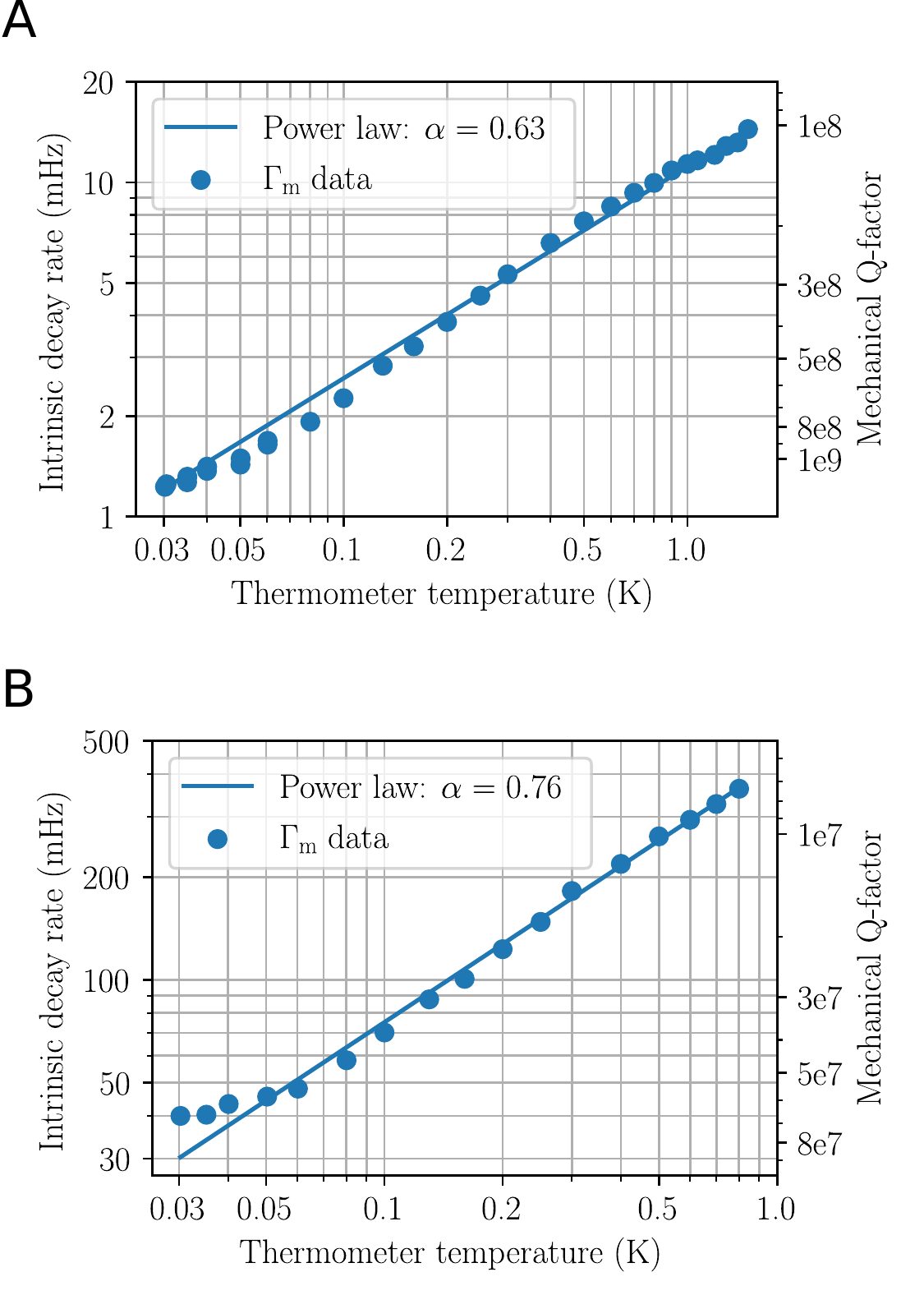}

\caption{Intrinsic mechanical decay rate as function of thermometer temperature. The power law fits indicate that coupling to mechanical TLS is the dominant loss mechanism for mechanical energy. We plot losses for two mechanical modes of the device: A) is the mode at \SI{1.486}{\MHz} discussed in the rest of the manuscript; B) is a mode at \SI{2.671}{\MHz}, located in a higher mechanical bandgap.}
    \label{fig:mechTLS}
\end{figure}

We measure the mechanics' intrinsic decay rate as function of thermometer temperature in Fig.~\ref{fig:mechTLS}. We fit the data to a power law of temperature $\propto (T/T_0)^\alpha$, with $T_0$ and arbitrary reference temperature and the exponent $\alpha$. For the two analysed mechanical modes, the power law relations with $\alpha = 0.63$ and $\alpha = 0.76$ are consistent with mechanical two level systems (TLS) coupling to the mode of interest and extracting mechanical energy from it\cite{Zhou2019a}.

We point out that since the mechanical mode in Fig.~\ref{fig:mechTLS}B at \SI{2.671}{\MHz} is located in the second mechanical bandgap, its spatial extend is less that for the \SI{1.486}{\MHz} mode, located in the first mechanical bandgap. Therefore the Al metallisation makes up a larger fraction of its out-of-plane displacement profile. The larger exponent $\alpha = 0.76$ could thus be attributed to the larger relative amount of Al in the mechanical mode.

\subsection{Cryogenic setup}

\begin{figure}[h!]
    \centering
    \includegraphics[width=1.\linewidth]{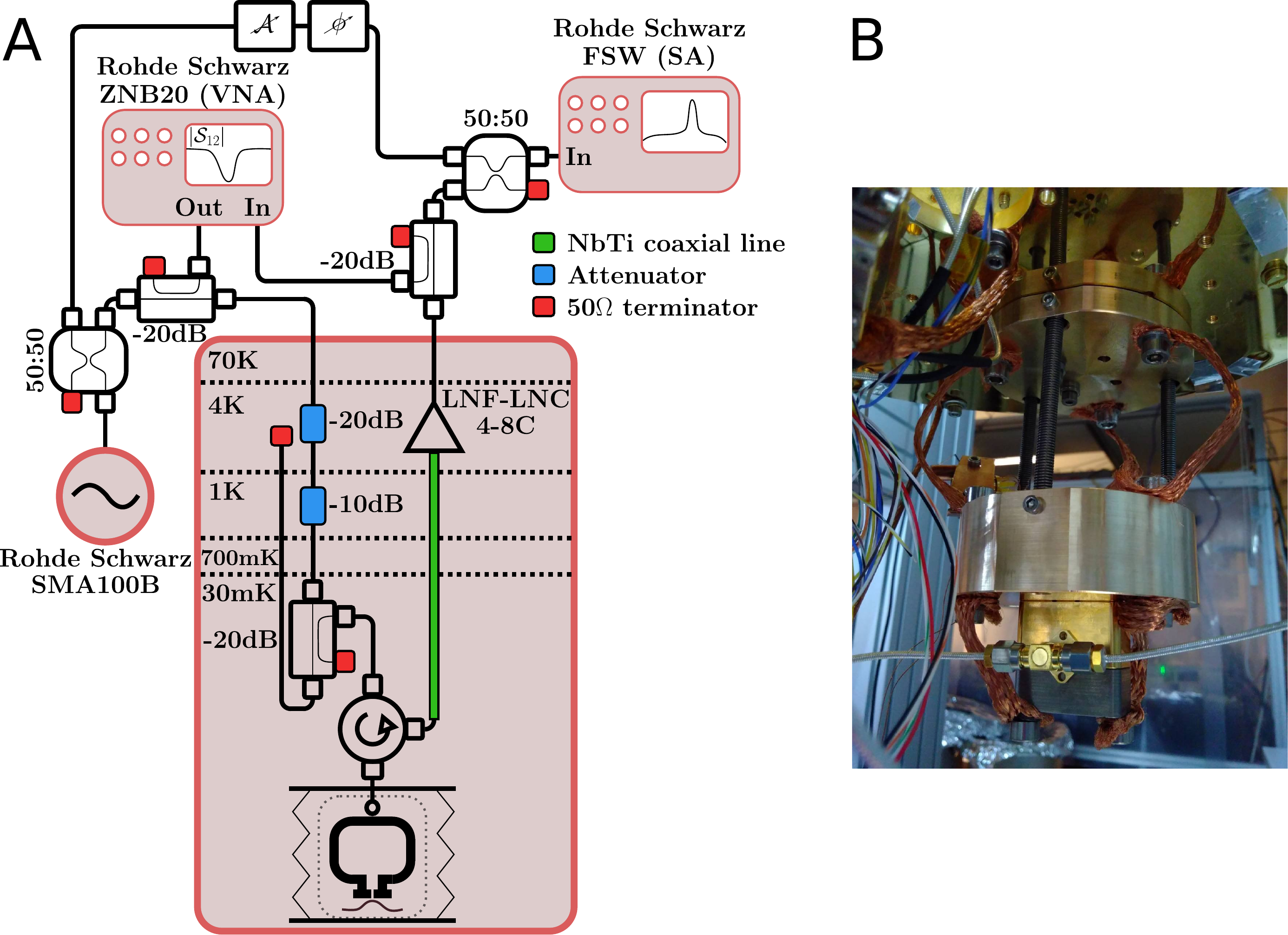}

\caption{A) Setup of the microwave equipment around the dilution refrigerator. B) Photograph of the single-stage mechanical damper mounted to the mixing chamber plate (top) of the dilution refrigerator to isolate from environmental vibrations, especially the pulse tube.}
    \label{fig:setup}
\end{figure}

In Fig.~\ref{fig:setup} A, we show a diagram of the microwave cable wiring used in this experiment. Fig.~\ref{fig:setup}B is a picture of the mechanical damper to which the electromechanical assembly is mounted. The damper consists of a oxygen-free high-conductivity copper block suspended from the mixing chamber plate by steele springs. This mass-on-a-spring setup acts as a mechanical low pass filter damping any environmental vibrations at the mechanical frequency. Each of the three springs has a spring constant of \SI{0.1}{\N/\mm} and the copper block has a mass of \SI{1.4}{\kg}, thus with the three springs in parallel we expect a low-pass filter corner frequency of \SI{\approx 0.5}{\Hz}, above which the suppression of vibration goes as $f^{-2}$.

\subsection{Calibration consistency check}
In order to verify that the thermal calibration, on which relies the core result of the manuscript, is correct, we performed two separate consistency checks:
\begin{itemize}
    \item Given the company specified gain and noise temperature of the HEMT at the operating frequency, we calibrated the spectrum data at low power from figure 4 of the main text, using the background noise as a reference. This translates, given the independent knowledge we have of the overcoupling ratio, the cooperativity and the mechanical linewidth, into a measured mechanical occupation, which is associated with a lower temperature than the extracted 80mK temperature bath. The ratio between those two temperatures is due to the attenuation between the sample and the HEMT amplifier. We found this attenuation to be approximately 6.0~dB. Separately, we summed the company specified losses of all the components, cables and connectors between the sample and the HEMT, and obtained 6.2~dB, with an estimated systematic error of $\pm1{.}5$ dB, given the limited applicability of the specifications to our low-temperature setting combined with the impossibility to accurately measure them at low temperature, the large number of components present and the uncertainties in connector losses for instance. 
    \item Alternatively, the thermal calibration provides a measurement of the gain between the sample and the analyzer, and combined with a transmission measurement of the whole setup, it allows to compute the attenuation between the source and the sample. We found this attenuation to be $66.5$ dB, and separately, summing again the company specified losses of all the components, we found $65.5$ dB of attenuation with an estimated systematic error of $\pm4$ dB.
\end{itemize}
Those two consistency checks strengthen our confidence in the thermal calibration presented in Fig 3 of the main text.

\newcommand{\etalchar}[1]{$^{#1}$}